# Velocity Saturation effect on Low Frequency Noise in short channel Single Layer Graphene FETs

Nikolaos Mavredakis*[a], Wei Wei[b], Emiliano Pallecchi[b], Dominique Vignaud[b], Henri Happy[b], Ramon Garcia Cortadella[c], Andrea Bonaccini Calia[c], Jose A. Garrido[c,d] and David Jiménez[a]

Graphene devices for analog and RF applications are prone to Low Frequency Noise (LFN) due to its upconversion to undesired phase noise at higher frequencies. Such applications demand the use of short channel graphene transistors that operate at high electric fields in order to ensure a high speed. Electric field is inversely proportional to device length and proportional to channel potential so it gets maximized as the drain voltage increases and the transistor's length shrinks. Under these conditions though, short channel effects like Velocity Saturation (VS) should be taken into account. Carrier number and mobility fluctuations have been proved to be the main sources that generate LFN in graphene devices. While their contribution to the bias dependence of LFN in long channels has been thoroughly investigated, the way in which VS phenomenon affects LFN in short channel devices under high drain voltage conditions has not been well understood. At low electric field operation, VS effect is negligible since carriers' velocity is far away from being saturated. Under these conditions, LFN can be precisely predicted by a recently established physics-based analytical model. The present paper goes a step furher and proposes a new model which deals with the contribution of VS effect on LFN under high electric field conditions. The implemented model is validated with novel experimental data, published for the first time, from CVD grown back-gated single layer graphene transistors operating at gigahertz frequencies. The model accurately captures the reduction of LFN especially near charge neutrality point because of the effect of VS mechanism. Moreover, an analytical expression for the effect of contact resistance on LFN is derived. This contact resistance contribution is experimentally shown to be dominant at higher gate voltages and is accurately described by the proposed model.

## Introduction

Extensive research has taken place the last decade after the discovery of graphene[1-2] due to its exceptional properties. Carrier mobilities up to $2.10^5 \, cm^2/V.s$ and saturation velocities of $4.10^7 \, cm/s$ led the scientific community to accept the challenge and take advantage of graphene in electronic applications by fabricating graphene transistors (GFETs)[3-4]. Despite the fact that graphene's zero bandgap is a deterrent for digital operation, the developments of GFETs for analog and RF applications is ongoing with very promising results. Frequency multipliers[5], voltage controlled oscillators[6] and THz detectors[7-9] are some examples of its electronic applications, while other applications of graphene are chemical-biological sensors[10-13] and optoelectronic devices[14]. The performance of all the above devices and circuits can be degraded by the effect of Low Frequency Noise (LFN) which can be up-converted to undesired phase noise in high frequency circuits and it can also affect the sensitivity of sensors[5-14]. In addition, LFN analysis can provide significant conclusions regarding the quality and reliability of graphene devices[15].

There are three main mechanisms that generate LFN in semiconductor devices: a) carrier number fluctuation ($\Delta N$), b) mobility fluctuation ($\Delta \mu$) and c) contact resistance ($R_c$) contribution ($\Delta R$). $\Delta N$ model[16] is based on trapping/detrapping mechanism where carriers can be captured and then emitted at border traps near the dielectric interface of a semiconductor[17]. A Random Telegraph Signal (RTS) in time domain which results in a Lorentzian spectrum is generated by each such trap. Supposing that these traps are uniformly distributed, then the superposition of these Lorentzians can cause an inversely proportional trend of the Power Spectral Density (PSD) of noise with frequency. For this reason the LFN is also known as 1/f (flicker) noise. In transistors with very small dimensions, the limited number of traps can lead to Lorentzian-shape PSDs, but this has not been yet observed in GFETs. $\Delta \mu$ model[18] is expressed by empirical Hooge formula and is considered to be caused by fluctuations of carrier mobility. Finally, $R_c$ can also influence LFN and this contribution can be very significant in GFETs at short channels where $R_c$ contribution cannot be neglected. There are many LFN models available in bibliography describing the above three effects for Metal-Oxide-Semiconductor FETs (MOSFETs)[19-22]. The same three mechanisms have also been found to be responsible for LFN in

[a] Departament d'Enginyeria Electrònica, Escola d'Enginyeria, Universitat Autònoma de Barcelona, Bellaterra 08193, Spain.
[b] Institute of electronics, Microelectronics and Nanotechnology, CNRS UMR8520, 59652 Villeneuve d'Ascq, France.
[c] Catalan Institute of Nanoscience and Nanotechnology (ICN2), CSIC, Barcelona Institute of Science and Technology, Campus UAB, Bellaterra, Barcelona, Spain.
[d] ICREA, Pg. Lluis Companys 23, 08010 Barcelona, Spain
* Email: nikolaos.mavredakis@uab.es

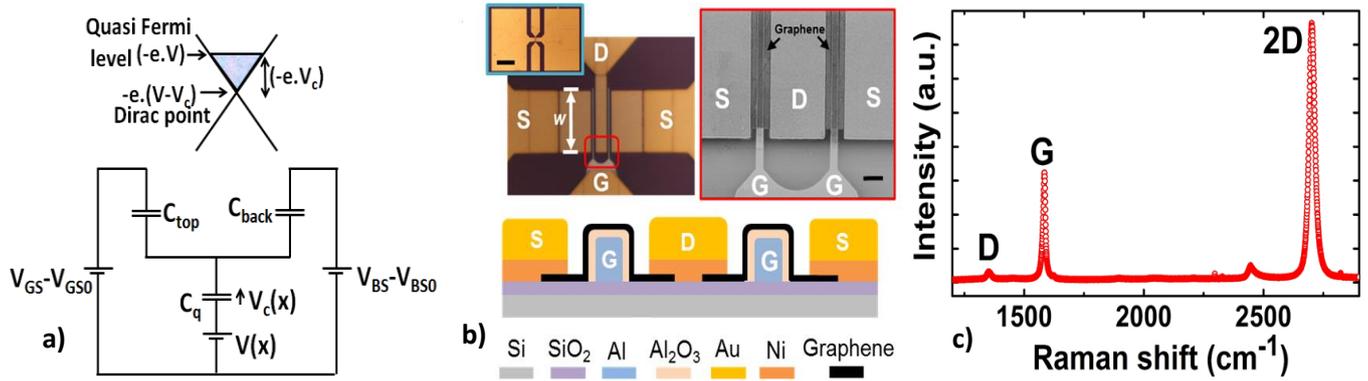

**Fig. 1** a) Energy dispersion relation of GFET (top) and its capacitive network (bottom) are shown with $C_q$: quantum capacitance, $C_{top}$, $C_{back}$: top and back oxide capacitances, $V_c(x)$: chemical potential, $V(x)$: quasi-Fermi channel potential, $V_{G(B)S}-V_{G(B)S0}$: top and back gate source voltage overdrives. (Top gate is not present in devices under test of (b) but is included in the capacitive network of (a) to support the generalizability of the model). b) GFET structure. Top left: Optical image of GFET. The channel width $W$ is $12\ \mu m$. The inset image shows the RF coplanar wave guide access. The scale bar is $60\ \mu m$. Top right: SEM image of our GFET with two-figures bottom gate structure. The scale bar is $1\ \mu m$. Bottom: Cross section schematic of our bottom gate structure. c) Raman spectra at $473\ nm$ laser excitation of graphene after device fabrication process.

GFETs[23-35]. $\Delta\mu$ effect is known to dominate in metals[18] and $\Delta N$ in semiconductors[17] where trapping/detrapping prevails, and since graphene can be considered a metal as well as a semiconductor, both of the above effects can contribute to its LFN. In fact, LFN nature in GFETs is strongly related to the number of layers since $\Delta N$ mechanism becomes more significant as this number is decreased while in multilayer GFETs $\Delta\mu$ is more important[24]. As it was mentioned before, $\Delta R$ can also play an important role because of the $R_c$ values in GFETs.

The most significant experimental characteristic of gate bias dependence of LFN in GFETs is the M-shape trend with a minimum close to charge neutrality point (CNP)[25-28, 34-35]. In a previous work[35], an analytical physics-based bias dependent model for long channel single-layer (SL) GFETs was proposed and successfully validated with experimental data. Both $\Delta N$ and $\Delta\mu$ models were shown to contribute to total LFN, especially near CNP, while the $R_c$ effect on LFN was observed at higher gate voltage regions but not analytically modeled. In addition, a strong relation between gate bias dependence of LFN and residual charge near CNP was shown; $\Delta N$ effect is responsible for M-shape behavior but as residual charge decreases, a $\Lambda$-shape trend can be observed[35]. Moreover, even a minor increase of drain voltage, was shown to slightly affect the homogeneity of the channel especially near CNP and this results in a small rise of LFN there. Those experiments were conducted at very small drain voltage values ($V_{DS}=20, 40, 60\ mV$) and thus, important phenomena that are significant at quite high electric fields such as Velocity Saturation (VS) could not be modeled.

Analytical modeling of LFN in short channel GFETs and the contribution of VS effect on it remain largely uninvestigated. In Si devices, VS effect causes a reduction of LFN at high electric fields[36] and this is also the case in GFETs as it will be shown in this work for the first time. VS effect is generated by optical phonon scattering mechanism and particularly for GFETs, saturation velocity $u_{sat}$ is usually approximated inversely proportional to chemical potential $V_c$[37-39]. While this relationship is acceptable away from CNP, it is not valid near the specific point where the chemical potential $V_c$ tends to 0 and thus $u_{sat}$ becomes very high, even higher than Fermi velocity $u_f(\sim 10^6\ m/s)$ which is the maximum velocity of carriers in graphene. Thus, a two branch model has been proposed[40-41] where a constant $u_{sat}$ value is considered for a quite low graphene net channel charge so that the GFET operates near CNP, while for higher values of charge, a more complicated energy dependent expression is used[40-41]. But if the aforementioned complicated model is used in LFN modeling for short channel GFETs, the equations become so complex, that it is generally impossible to find an analytical solution. That is why, an inversely proportional relation between $V_c$ and $u_{sat}$ is considered away from CNP.

The fundamental scope of this work is the extension of the model proposed in ref. 35 in order to include the VS effect on LFN. Furthermore, a simple analytical expression for $\Delta R$ contribution taken from Si devices[42] is proposed. As described thoroughly in ref. 35, the LFN model is implemented based on the assumption that the GFETs' channel is divided into infinitesimal slices, each of which corresponds to a local noise source[19, 22, 36, 42]. All these local noise sources can be considered uncorrelated and thus, the sum of their PSDs results in the total LFN[42]. In simpler words, by integrating all the local noise contributors along the device channel, the PSD of each LFN mechanism can be calculated. As it will be shown, these integrals can be solved analytically, similarly to ref. 35, based on a chemical potential based compact model[43-45] and thus, the LFN model can be easily implemented in Verilog-A and integrated in circuit simulators. The equivalent capacitive circuit of this model[43-45] is shown in Fig. 1a. Graphene charge $Q_{gr}$ is stored in the quantum capacitance ($C_q$); the chemical potential $V_c(x)$ represents the voltage drop across $C_q$ at position $x$. $V_c(x)$ is defined as the difference between the potential at quasi-Fermi level and the potential at the CNP, as shown in the energy dispersion relation scheme of graphene in Fig. 1a where $V_c(0)=V_{cs}$ at the source end ($x=0$) and $V_c(L)=V_{cd}$ at the drain end ($x=L$).

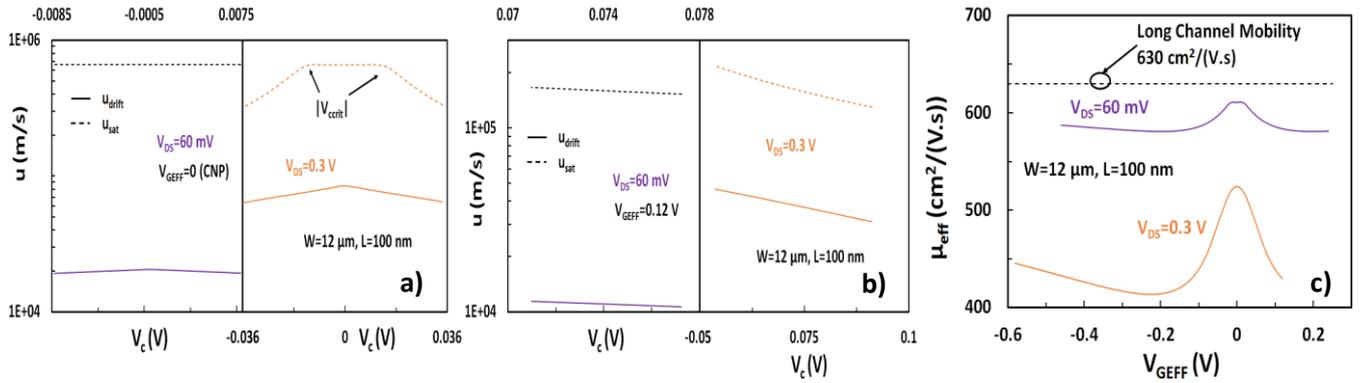

**Fig. 2** Drift velocity $u_{drift}$ vs. chemical potential $V_c$ for $W/L=12 \mu m/100 nm$ GFET with a) $V_{GEFF}=0 V$ *(CNP)* and b) $V_{GEFF}=0.12 V$ at $V_{DS}=60 mV$ (left subplot), *0.3 V* (right subplot). $u_{sat}$ is also shown with dashed lines. c) Effective mobility $\mu_{eff}$ vs back gate voltage overdrive $V_{GEFF}$ for $W/L=12 \mu m/100 nm$ GFET at low and high $V_{DS}$ of *60 mV* and *0.3 V*. Constant long channel mobility (model parameter $\mu$) is also shown with dashed line.

$V_{GS}$-$V_{GSO}$, $V_{BS}$-$V_{BSO}$ are the top and back gate source voltage overdrives while $C_{top}$ and $C_{back}$ are the top and back gate capacitances, respectively. The sum of top and back gate capacitances is defined as $C=C_{top}+C_{back}$. The quasi-Fermi potential $V(x)$ is the voltage drop in the graphene channel at position $x$, which is equal to zero at the source end ($x=0$) and equal to $V_{DS}$ at the drain end ($x=L$). A good agreement between drain current data and the above model[43-45] is crucial for the good performance of LFN model, since IV (current-voltage) quantities are used in LFN expressions.

The extracted LFN model is validated with experimental data from bottom-gated SL GFETs where graphene grown by CVD on a copper foil was used[46-49]. The optical and SEM images as well as the schematic of the graphene device are shown in Fig. 1b. Details on the fabrication process can be found in Experimental data section. The quality of graphene after transfer process was verified by performing Raman characterization. Fig. 1c shows a typical Raman spectra of graphene on the substrate ($SiO_2/Si$) after the device fabrication process. The 2D peak at 2703 cm$^{-1}$ is fitted with a single Lorentzian component with a full width at half-maximum (FWHM) of 26 cm$^{-1}$. The G peak locates at 1584 cm$^{-1}$ and has a FWHM (G) of 12 cm$^{-1}$.The ratio of the 2D and G peak integrated intensities stands around 2.5, which indicates single layer graphene. The intensity ratio of the D and G peak is very low, ~0.1, which suggests that a very low defect density is present in our fabricated devices.

## Results and Discussion

The drain current model proposed in ref. 43- 45 and on which the LFN analysis of the present work is based, assumes a drift-diffusion carrier transport with a soft VS model[37-41]. In more detail:

$$I_D=WQ_{gr}\mu_{eff}E, \ \mu_{eff}=\frac{\mu}{1+|E_x|/E_c} \quad (1)$$

where $W$ is the width of GFET, $\mu$ is the low field carrier mobility, $E$ is the electric field and $\mu_{eff}$ is the effective carrier mobility which represents the degradation of mobility $\mu$ at high electric fields and depends on the ratio of longitudinal electrical field $E_x$ and the critical field $E_c$. $E_c$ is the value of electric field $E_x$ above which the carriers' velocity saturates. After a more detailed analysis (see eqn (A1-A2) in ESI A), eqn (2) (bottom of the page) is derived which represents the change of integral variable from $x$ to $V_c$ and it will be proved to be very significant for the derivation of the IV and LFN analytical expressions; VS effect is considered through saturation velocity term $u_{sat}$. Then by integrating eqn (2) along the device channel from Source (S) to Drain (D):

$$I_D=\frac{-\int_{V_s}^{V_d}WQ_{gr}\mu dV}{L+\int_{V_s}^{V_d}\frac{\mu}{u_{sat}}\left(\frac{C_q+2C}{C_q+C}\right)dV}=\frac{W\int_{V_{cd}}^{V_{cs}}Q_{gr}\frac{C_q+C}{C}dV_c}{L+\int_{V_{cd}}^{V_{cs}}\frac{\mu}{u_{sat}}\left(\frac{C_q+2C}{C}\right)dV_c} \quad (3)$$

The denominator of the eqn (3) expresses an effective channel length $L_{eff}$ which accounts for the degradation of $I_D$ because of VS effect (see see eqn (A6-A9) in ESI B). The sign of $V_{DS}$ determines the sign of the electric field $E$ and consequently, the sign of the longitudinal electrical field $E_x$ as described thoroughly in ESI A, B. Regarding the value of $u_{sat}$, a two-branch model is used, as mentioned before:

$$u_{sat}=\begin{cases} \frac{2u_f}{\pi}=S, \ V_c \leq V_{ccrit} \\ \frac{\Omega}{\sqrt{\frac{\pi Q_{gr}}{2e}}}=\frac{\Omega}{\sqrt{\frac{\pi k\left(V_c^2+a/k\right)}{2e}}}=\frac{N}{\sqrt{V_c^2+a/k}}, \ N=\frac{h\Omega u_f}{e}, \ V_c > V_{ccrit} \end{cases} \quad (4)$$

The analytical expressions for coefficient $k$, Fermi velocity $u_f$, graphene charge $Q_{gr}$, bias dependent term $g(V_c)$ and residual charge related term $\alpha$ are given in ESI A, $h\Omega$ is the phonon energy and $e$ is the electron charge. $Q_{crit}$ is the critical value of graphene net charge above which $u_{sat}$ is considered inversely proportional to $V_c$ as shown in bottom branch of eqn (4) while it is constant below $Q_{crit}$ as it is shown in upper branch of eqn (4). $Q_{crit}=e\Omega^2/(2\pi u_f^2)$ and from there $V_{ccrit}$ can also be calculated.

All plots of Fig. 2 are for a GFET with $L=100 nm$. In Fig. 2a and 2b, the drift velocity $u_{drift}$ and the saturation velocity $u_{sat}$ are shown vs. the chemical potential $V_c$ at the CNP and away from it respectively. In each graph, $u_{drift}$ at low (left subplot) and high

$$\frac{dx}{dV} = \frac{-Q_{gr}\,2L_{eff}}{kg_{vc}} + \frac{\mu}{\upsilon_{sat}}\frac{C_q+2C}{C_q+C} \Rightarrow \frac{dx}{dV_c} = \frac{-Q_{gr}\,2L_{eff}}{kg_{vc}}\left(\frac{C_q+C}{C}\right) + \frac{\mu}{\upsilon_{sat}}\left(\frac{C_q+2C}{C}\right) \quad if \quad V_{ds} \geq 0$$

$$\frac{dx}{dV} = \frac{-Q_{gr}\,2L_{eff}}{kg_{vc}} - \frac{\mu}{\upsilon_{sat}}\frac{C_q+2C}{C_q+C} \Rightarrow \frac{dx}{dV_c} = \frac{-Q_{gr}\,2L_{eff}}{kg_{vc}}\left(\frac{C_q+C}{C}\right) - \frac{\mu}{\upsilon_{sat}}\left(\frac{C_q+2C}{C}\right) \quad if \quad V_{ds} \leq 0$$

(2)

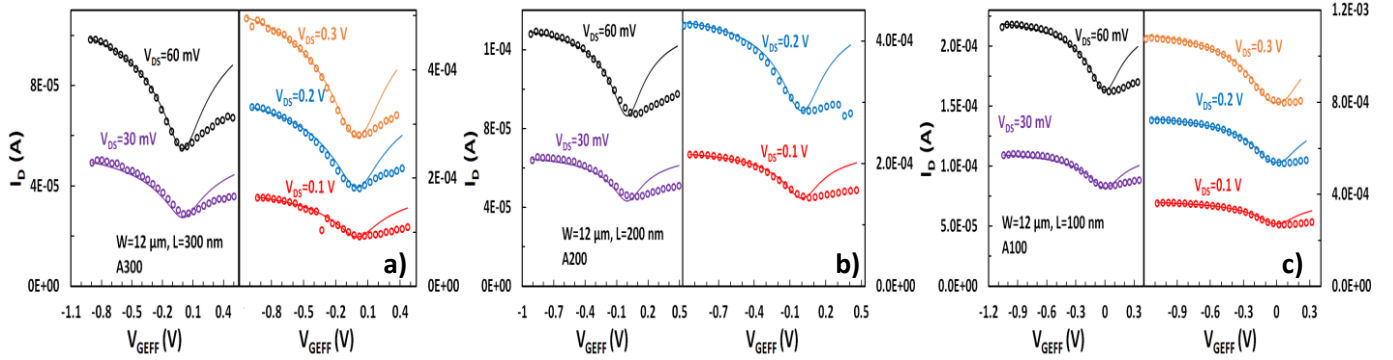

**Fig. 3** Drain current $I_D$ vs. back gate voltage overdrive $V_{GEFF}$, for GFETs with $W=12\ \mu m$ and a) $L=300\ nm$ (A300), b) $L=200\ nm$ (A200) and c) $L=100\ nm$ (A100) at low (left subplot) and high (right subplot) available $V_{DS}$ values ($V_{DS}=30\ mV,\ 60\ mV,\ 0.1\ V,\ 0.2\ V$ and $0.3\ V$). markers: measured, solid lines: model.

(right subplot) drain voltage are shown ($V_{DS}=60\ mV$ and $0.3\ V$) as a solid line; $u_{sat}$ is also shown with dashed lines. At low drain voltage, $u_{drift}$ is much smaller than $u_{sat}$ and thus, VS effect is negligible. This is the case both at and away from the CNP. On the contrary, at higher drain voltage, $u_{drift}$ is still smaller but comparable to $u_{sat}$ which means that VS effect starts to become significant for every gate voltage regime. This is also shown in terms of electric field $E_x$ in ESI C (Fig. S1). In Fig. 2c, effective mobility $\mu_{eff}$ is shown vs. effective gate voltage $V_{GEFF}$ (back-gate voltage overdrive) again for both drain voltage values mentioned before. As it was expected at low drain voltage, $\mu_{eff}$ is quite close to long channel mobility $\mu$, while for higher drain voltage, VS effect causes a significant degradation of $\mu_{eff}$. In accordance with the $u_{sat}$ model described in eqn (4), effective mobility is shown to get maximized at CNP. In more detail and as it is shown in Fig. 2a and 2b, $u_{sat}$ becomes maximum at CNP and consequently $E_c$, which is proportional to $u_{sat}$, is also maximized (see eqn (A1) in ESI A) while the ratio $E_x/E_c$ gets minimum (see Fig. S1 in ESI C). As a result from eqn (1), $\mu_{eff}$ is maximum at CNP.

IV and LFN data were measured for six different GFETs with $W=12\ \mu m$ and for three available channel lengths; $L=300\ nm$ for A300, B300 devices, $L=200\ nm$ for A200, B200 devices and $L=100\ nm$ for A100, B100 devices (see Experimental Data section for more details on fabrication and measurements). The back gate voltage was swept from $V_G=0$ to $1.4\ V$ with a step of $50\ mV$ for transfer characteristics while for LFN spectra the sweep was from $V_G=0.6$ to $1.3\ V$ with a step of $50\ mV$, covering regions both away and near CNP. The measured frequency range was from $1\ Hz$ to $1\ kHz$. Moreover, five different drain voltage values were recorded for both IV and LFN setups ($V_{DS}=30m,\ 60m,\ 0.1,\ 0.2$ and $0.3\ V$) in order to cover the low and high electric field region which is crucial for studying VS effect. In a few cases, some drain voltages were omitted since the IV as well as LFN data were completely out of order, probably because of leakages or possible break down of the devices at higher electric fields. In more detail, for A300, A100 and B100 GFETs all five drain voltages were measured, for B200, A200 GFETs, $V_{DS}=0.3\ V$ is missing while for B300 GFET, $V_{DS}=30\ mV$ is missing. Fig. 3 presents the transfer characteristics of a) A300, b) A200 and c) A100 GFETs at all available drain voltages. The compact model reported in ref. 43-45 was used for simulating drain current and the fitting with experimental data is of high quality both near and away CNP at p-type region. There is an asymmetry in drain current data at higher gate voltages since they are lower in n-type than p-type region and this can be explained either by a different mobility at the two operating regimes or by a parasitic p-n (n-n) diode which is formed between the channel and the contact when the device is biased in the p(n)-region giving rise to different contact resistances[50] .The model is symmetric so it can provide identical behaviour at p- and n-type regimes away from CNP. The best fit was achieved at the whole p-type region, near CNP and up to a value of $V_{GEFF}\approx0.15\ V$ in n-type regime. We focused our attention on this region which presents the highest transconductance, a crucial figure of merit for RF applications. Table 1 presents the parameter set of the IV model which includes the long channel carrier mobility ($\mu$), the back gate capacitance ($C_{back}$), the flat band back gate voltage ($V_{BSO}$), the contact resistance ($R_c$), the inhomogeneity of the electrostatic potential ($\Delta$) which is related to the residual charge density $\rho_0$[43-45] and the phonon energy ($h\Omega$) which is related to VS effect. One parameter set is used for all bias conditions at each GFET; all the above parameters except $h\Omega$ are extracted for low drain voltages, while $h\Omega$ is extracted for the higher ones. Fig. 4a shows the equivalent noise subcircuit[35, 42] where a random local current noise source $\delta I_n$ with a PSD $S_{\delta I}{}^2_n$ is used to model the local fluctuations. This noise subcircuit and its operating principles is analysed thoroughly in ref. 35. Fig. 4b and 4c illustrate the LFN

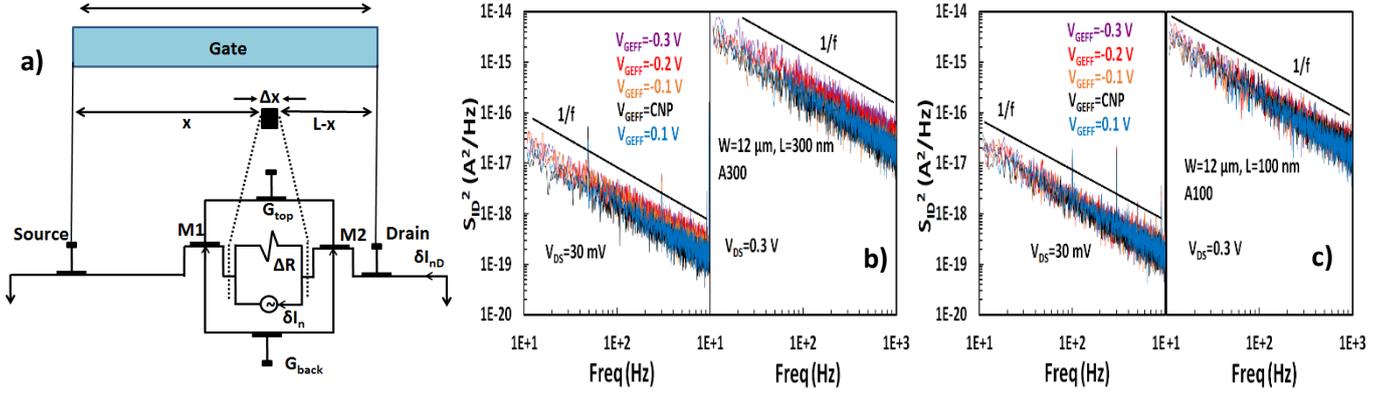

**Fig. 4** a) Equivalent noise subcircuit with a local current noise source. Relative power spectral density of drain current noise $S_{iD}$ for b) $W/L=12\ \mu m/300\ nm$ A300 GFET and c) $W/L=12\ \mu m/100\ nm$ A100 GFET with back gate voltage overdrive ($V_{GEFF} = -0.3, -0.2, -0.1, 0, 0.1\ V$) at low $V_{DS}= 30\ mV$ (left subplot) and high $V_{DS}= 0.3\ V$ (right subplot). The solid line corresponds to a 1/f slope.

$$\Delta NA = \frac{A1}{2k\left(\alpha k+C^2\right)g_{vc}}\left[-2\sqrt{\alpha k}\,C\arctan\left(\sqrt{\frac{k}{\alpha}}V_c\right)\pm\alpha k\ln\left(\alpha+kV_c^2\right)\pm2C^2\ln\left(C\pm kV_c\right)\right]_{V_{cd}}^{V_{cs}} \tag{5}$$

$$\Delta NB\big|_{V_c|\leq V_{ccrit}}=\frac{B1}{2\left(\alpha k+C^2\right)^3 S}\left[\begin{array}{c}\mp\dfrac{2C^3k\left(C^2+ak\right)}{C+kV_c}\mp\dfrac{k\left(C^2+ak\right)\left(3aC^2+a^2k\pm2C^3V_c\right)}{\alpha+kV_c^2}+\dfrac{2C^3\sqrt{k}\left(C^2-3\alpha k\right)}{\sqrt{\alpha}}\arctan\left(\sqrt{\dfrac{k}{\alpha}}V_c\right)\\[3mm]\pm k\left(-3C^4+aC^2k\right)\ln\left[\dfrac{a+kV_c^2}{\left(C\pm kV_c\right)^2}\right]\end{array}\right]_{V_{cd}}^{V_{cs}} \tag{6a}$$

$$\Delta NB\big|_{V_c|\geq V_{ccrit}}=\frac{B1}{N}\left[\dfrac{\mp k\sqrt{V_c^2+\dfrac{\alpha}{k}}\left(a^2k\left(\pm C+kV_c\right)+C^3V_c\left(2C\pm3kV_c\right)+aC^2\left(\pm4C+3kV_c\right)\right)}{\left(C^2+ak\right)^2\left(\pm C+kV_c\right)\left(\alpha+kV_c^2\right)}\pm\dfrac{C^2\left(2C^2-ak\right)\ln\left[\dfrac{C\pm kV_c}{\alpha\mp CV_c+\sqrt{C^2+ak}\sqrt{V_c^2+\dfrac{\alpha}{k}}}\right]}{\left(C^2+ak\right)^{5/2}}\right]_{V_{cd}}^{V_{cs}} \tag{6b}$$

$$\Delta\mu A=\frac{A2}{g_{vc}}\left[CV_c\pm\frac{kV_c^2}{2}\right]_{V_{cd}}^{V_{cs}} \tag{7} \qquad\qquad \Delta\mu B\big|_{V_c|\leq V_{ccrit}}=\frac{B2k}{S}\left[\frac{2C}{\sqrt{\alpha k}}\arctan\left(\sqrt{\frac{k}{\alpha}}V_c\right)\pm0.5\ln\left(a+kV_c^2\right)\right]_{V_{cd}}^{V_{cs}} \tag{8a}$$

$$\Delta\mu B\big|_{V_c|\geq V_{ccrit}}=\frac{B2}{N}\left[\pm k\sqrt{V_c^2+\frac{\alpha}{k}}+2C\ln\left(V_c+\sqrt{V_c^2+\frac{\alpha}{k}}\right)\right]_{V_{cd}}^{V_{cs}} \tag{8b} \qquad \Delta N=\Delta NA-\Delta NB,\quad\Delta\mu=\Delta\mu A-\Delta\mu B \tag{9}$$

$$\Delta R=\frac{g_{ms}^2+g_{md}^2}{\left[1+\dfrac{R_c}{2}\left(g_{ms}+g_{md}\right)\right]^2}S_{\Delta R^2},\quad g_{ms,d}=\frac{\mu Wk}{2L_{eff}}\frac{C_{back}}{C}V_{cs,d}^2 \tag{10}$$

* $\pm,\mp$ : Top sign refers to $V_c>0$ and bottom sign to $V_c<0$.

* Eqns (6a, 6b, 8a, 8b) are defined in cases where limits of integration have the same sign and the absolute value of both of them is below (eqns (6a, 8a)) or above (eqns (6b, 8b)) $V_{ccrit}$. See main text below and ESI E for better understanding.

* Eqns (6a, 8a) correspond to the integral: "LFN Near CNP" while eqns (6b, 8b) correspond to the integral "LFN Away CNP" which are defined in ESI E.

spectra of A300 and A100 GFETs respectively with a slope close to 1/f, at five different effective gate potentials ($V_{GEFF}$=-0.3, -0.2, -0.1, $V_{CNP}$, 0.1 V). Low and high drain voltages are shown ($V_{DS}$= 30 mV, 0.3 V) at left and right subplots respectively. The 1/f trend of LFN is apparent in all cases.

The derivation of $\Delta N$ and $\Delta \mu$ models is thoroughly analysed in ref. 35 by considering a linear dependence of the quantum capacitance $C_q$ and the chemical potential $V_c$ ($C_q$=$k \cdot |V_c|$) [35, 43]. $N_T$, which is the dielectric volumetric trap density per unit energy (in $eV^{-1}cm^{-3}$), is used as a first model parameter related to $\Delta N$ effect[35] while $\alpha_H$, which is the unitless Hooge parameter, is used as a second model parameter related to $\Delta \mu$ effect[35]. For both $\Delta N$ and $\Delta \mu$ cases, the total PSD of normalized drain current noise divided by squared drain current at 1 Hz is calculated by considering the integral from S to D (see eqn (A15, A22) for $\Delta N$ and $\Delta \mu$, respectively in ESI D). Then by applying eqn (2), the integral variable of LFN changes from x to $V_c$ (see eqn (A16, A23) for $\Delta N$ and $\Delta \mu$, respectively in ESI D). The latter is essential since the IV model is a chemical potential based model[35, 43-45]. Since eqn (2) has two terms on the right hand side, this results in two $\Delta N$ related terms, $\Delta NA$ and $\Delta NB$, and two $\Delta \mu$ related terms, $\Delta \mu A$ and $\Delta \mu B$, which are derived in the format of integrals before being solved analytically. (see eqns (A19-A20, A26-A27) of ESI D). VS phenomenon contributes to LFN both through $L_{eff}$, which is contained in constants A1, B1, A2, B2 which are defined in ESI D, and through second right hand term of eqn (2). $\Delta NA$ and $\Delta \mu A$, which come from the first right hand term of eqn (2), do not include $u_{sat}$ and in fact are the same with the $\Delta N$-$\Delta \mu$ LFN terms extracted in ref. 35 with the only difference of presenting $L_{eff}$ instead of L. $\Delta NB$ and $\Delta \mu B$ on the other hand, include $u_{sat}$ since they come from the second right hand term of eqn (2) (see eqns (A29 and A31) for $\Delta NB$ and eqns (A30 and A32) for $\Delta \mu B$ of ESI D) and they represent the main correction to LFN that is proposed in this work along with the contribution of $L_{eff}$. The behaviour of all the contributing terms to LFN locally in the channel is shown in Fig. S2 of ESI D. The integrals of $\Delta NA$, $\Delta NB$, $\Delta \mu A$ and $\Delta \mu B$ terms are then solved analytically in eqns (5-8) for the case of a positive drain voltage. Effective length $L_{eff}$ can be solved analytically (see eqns (A10-A11) in ESI B). Eqn (9) shows that the VS induced LFN

($\Delta NB$, $\Delta \mu B$) is subtracted by the long channel LFN ($\Delta NA$, $\Delta \mu A$) for both $\Delta N$ and $\Delta \mu$ noise contributions which agrees with the findings in MOSFETs[36]. Even for a negative drain voltage the VS effect reduces LFN. More specifically, $\Delta NA$, $\Delta \mu A$ terms do not change sign with negative drain voltage since they are derived from the first term of the right hand of eqn (2) where its sign is not affected by drain voltage polarity. This can be proved since the integration of $\Delta NA$, $\Delta \mu A$ in eqns (5, 7) from $V_{cs}$ to $V_{cd}$ gives negative results but $g_{VC}$ is also negative and thus, $\Delta NA$, $\Delta \mu A$ LFN terms are positive as in the case of positive drain voltage. On the other hand, $\Delta NB$, $\Delta \mu B$ terms are derived from the second term of the right hand of eqn (2) and thus, they change their signs for negative drain voltage. In this case, eqns (6, 8) remain identical with just calculating the integrals from $V_{cd}$ to $V_{cs}$, which results in positive solution. Thus, eqn (9) always results in reduction of $\Delta N$, $\Delta \mu$ terms regardless of the polarity of drain voltage. It is very crucial to notice that during the integration process (see eqns (A19, A26, and A29-A32) of ESI D) that results in the analytical expressions of eqns (5-8), different cases should be considered depending both on the signs of $V_{cs}$, $V_{cd}$ and if their absolute value is lower or higher than $V_{ccrit}$ since different equations are valid in each such region. In more detail, there are four separate regions where $V_c$ might be; a) lower than negative $V_{ccrit}$, b) higher than negative $V_{ccrit}$ but lower than $V_c$=0 (CNP), c) higher than $V_c$=0 (CNP) but lower than positive $V_{ccrit}$ and d) higher than positive $V_{ccrit}$. If both $V_{cs}$, $V_{cd}$ belong to the same region then the eqns (5-8) are solved from $V_{cs}$ to $V_{cd}$. On the other hand, if $V_{cs}$, $V_{cd}$ belong to different regions then the integrals in eqns (5-8) should be split into as many sub-integrals as required, corresponding to the four regions mentioned above, to ensure that the limits of integration of each one of the new sub-integrals have the same polarity and their absolute values are both higher or lower than $V_{ccrit}$. This happens since eqns (5-8) are valid and can be solved only in each such region. This process is described in detail in eqns (A33- A34) of ESI E where the sub-integrals are added when is needed in order to take the total solution.

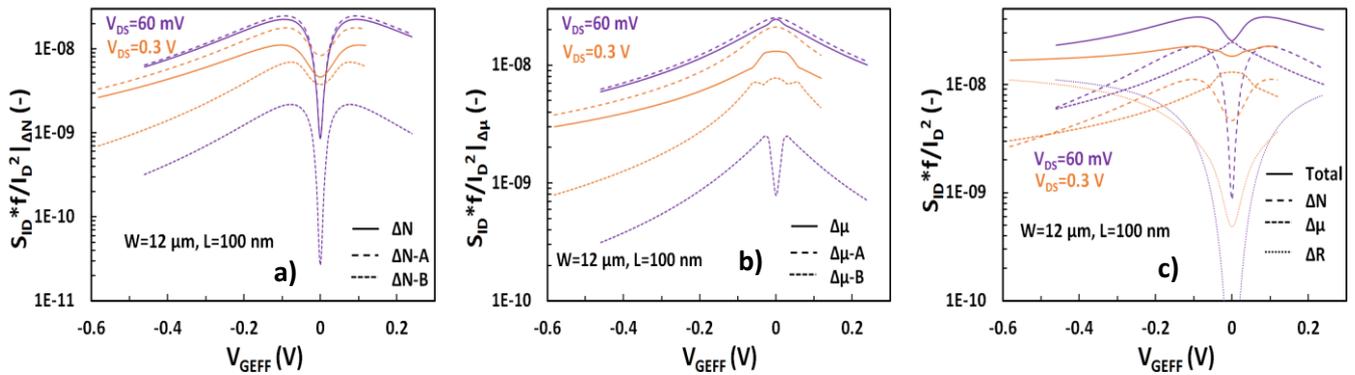

**Fig. 5** Output current noise divided by squared drain current $S_{ID}/I_D^2$, referred to 1 Hz, vs. back gate voltage overdrive $V_{GEFF}$, for a) $\Delta N$ effect, b) $\Delta \mu$ effect and c) all noise contributions for W/L=12 μm/100 nm GFET. Dashed lines represent $\Delta NA$, $\Delta NB$ contributors of $\Delta N$ effect in (a), $\Delta \mu A$, $\Delta \mu B$ contributors of $\Delta \mu$ effect in (b) and total $\Delta N$ and $\Delta \mu$ noise mechanisms in (c). $\Delta R$ contributor is also shown with dotted lines in (c).

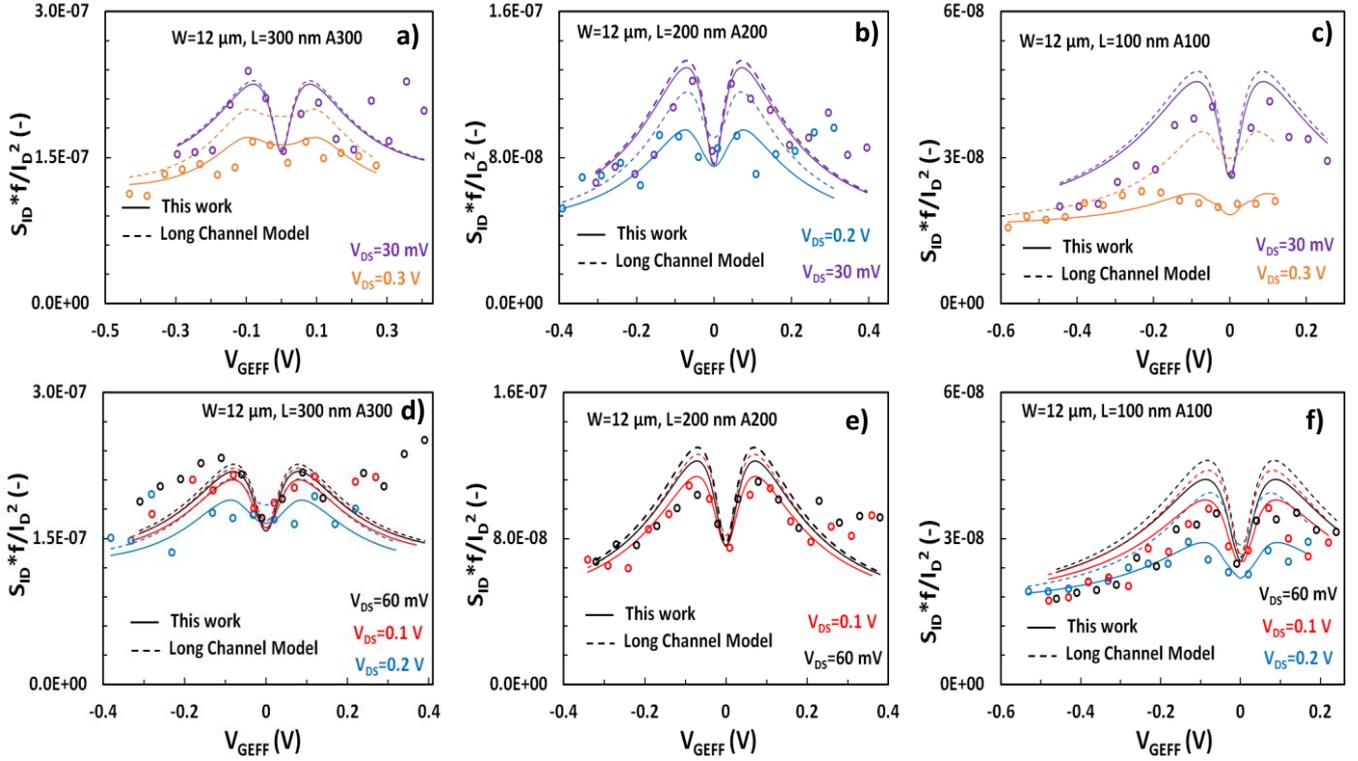

**Fig. 6** Normalized drain current noise divided by squared drain current referred to *1 Hz*, $S_{ID}f/I_D^2$, vs. back gate voltage overdrive $V_{GEFF}$, for GFETs with *W=12 μm* and a, d) *L=300 nm* (A300), b, e) *L=200 nm* (A200) and c, f) *L=100 nm* (A100): Upper plots show the highest and lowest available $V_{DS}$ value depending on the GFET. (A300: $V_{DS}$=30 mV, 0.3 V, A200: $V_{DS}$=30 mV, 0.2 V, A100: $V_{DS}$=30 mV, 0.3 V) while down plots show the rest of available $V_{DS}$ values (A300: $V_{DS}$=60 mV, 0.1 V, 0.2 V, A200: $V_{DS}$=60 mV, 0.1 V, A100: $V_{DS}$=60 mV, 0.1 V, 0.2 V) markers: measured, solid lines: model, dashed lines: long channel model from ref. 35.

Table 1. IV and LFN Noise model Parameters

| Parameter | Units | A300 (L=300 nm) | B300 (L=300 nm) | A200 (L=200 nm) | B200 (L=200 nm) | A100 (L=100 nm) | B100 (L=100 nm) |
|---|---|---|---|---|---|---|---|
| $\mu$ | cm²/(V·s) | 300 | 300 | 670 | 670 | 630 | 300 |
| $C_{back}$ | μF/cm² | 1,35 | 1,35 | 1,35 | 1,35 | 1,35 | 1,35 |
| $V_{BSO}$ | V | 0,88 | 0,92 | 0,89 | 0,89 | 1,03 | 0,9 |
| $\Delta$ | eV | 0.105 | 0.116 | 0.082 | 0.082 | 0.097 | 0.105 |
| $h\Omega$ | eV | 0.016 | 0.016 | 0.017 | 0.017 | 0.018 | 0.017 |
| $R_c/2=R_{S,D}$ | Ω | 260 | 198 | 219 | 223 | 131 | 176 |
| $N_T$ | eV⁻¹cm⁻³ | 8·10¹⁹ | 7·10¹⁹ | 2.1·10¹⁹ | 2.3·10¹⁹ | 5.5·10¹⁸ | 1.5·10¹⁹ |
| $\alpha_H$ | - | 5·10⁻³ | 3·10⁻³ | 1.1·10⁻³ | 1.5·10⁻³ | 2.5·10⁻⁴ | 9.5·10⁻⁴ |
| $S_{\Delta R^2}$ | Ω²/Hz | 2·10⁻² | 7·10⁻³ | 3.4·10⁻³ | 9·10⁻³ | 5·10⁻⁴ | 4.5·10⁻³ |

In the present work, a simple model for $\Delta R$ contribution is also derived taken from Si devices[42] since the effect of $R_c$ on LFN is significant especially at higher-gate voltages as it will be shown later. $\Delta R$ model is described in eqn (10) where $S_{\Delta R^2}$ expressed in $\Omega^2/Hz$, is the third parameter of the proposed LFN model and $g_{ms}$, $g_{md}$ are the source and drain transconductances[45]. In order to calculate the total LFN, the three different contributions have to be added as:

$$\frac{S_{I_D}}{I_D^2} = \frac{S_{I_D}}{I_D^2}\bigg|_{\Delta N} + \frac{S_{I_D}}{I_D^2}\bigg|_{\Delta \mu} + \frac{S_{I_D}}{I_D^2}\bigg|_{\Delta R} \tag{11}$$

The behaviour of all compact expressions of LFN related terms of eqns (5-10) is analysed in Fig. 5. Normalized drain current LFN divided by squared drain current and referred to *1 Hz*, $S_{ID}f/I_D^2$ is shown vs. effective gate voltage $V_{GEFF}$ for a low and a high drain voltage value ($V_{DS}$=60 mV, 0.3 V). In Fig. 5a $\Delta NA$, $\Delta NB$ and $\Delta N$ terms of carrier number fluctuation mechanism are presented. $\Delta N$ effect is responsible for M-shape of LFN[35] and this is also confirmed in Fig. 5a. Apart from $\Delta NA$ long channel term, $\Delta NB$ also follows an M-shape trend. For low drain voltage, $\Delta NB$ is negligible and $\Delta NA$ coincides with $\Delta N$. This is totally acceptable since at this region, VS effect does not contribute at all to drain current as well as to LFN. On the other hand, for higher drain voltage, the contribution of VS effect is apparent. The carrier number fluctuation term $\Delta N$ which is calculated by the subtraction of $\Delta NA$ and $\Delta NB$ as it is shown in eqn (9), is significantly lower than $\Delta NA$ which in fact represents the long channel case. Similar conclusions can be extracted in Fig. 5b where the $\Delta \mu A$, $\Delta \mu B$ and $\Delta \mu$ terms of mobility fluctuation effect are shown. Again, for low drain voltage, $\Delta \mu B$ is much lower than $\Delta \mu A$ and as a consequence $\Delta \mu A$ dominates. For high drain voltage though, the difference $\Delta \mu A$- $\Delta \mu B$, which results in $\Delta \mu$ according to eqn (9) is significant. VS effect is responsible for the reduction of both $\Delta N$ and $\Delta \mu$ LFN mechanisms and this is the case at higher electric field regime. Even the long channel terms $\Delta NA$ and $\Delta \mu A$ are lower in $V_{DS}=0.3\ V$ than $V_{DS}=60\ mV$ and this can be explained by the presence of $L_{eff}$ in denominators of eqns $A1$, $A2$ of eqns (5, 7) respectively (See ESI D, eqns (A19, A26)). Finally at Fig. 5c all three contributions of normalized LFN ($\Delta N$, $\Delta \mu$ and $\Delta R$) are shown. $\Delta N$ effect is responsible for the M-shape which becomes less deeper at higher drain voltage due to non-homogeneous channel condition[35]. $\Delta \mu$ effect contributes near CNP[35] for both high and low drain voltages while $\Delta R$ is negligible near CNP but has a strong impact at higher gate voltages where $R_c$ is also dominant. Away from CNP, $\Delta R$ contribution is similar for both drain voltage levels while at CNP, where $\Delta R$ is anyway negligible, its level is much lower at low drain voltage value. The later can by justified in terms of eqn (10) since $V_{cs,d}$ tend to 0 at low drain voltage and near CNP operating conditions.

The proposed analytical LFN model is validated with experimental data from devices under test[46-49]. In Fig. 6 we show the normalized total LFN $S_{ID}f/I_D^2$ vs. effective gate voltage $V_{GEFF}$ for the devices whose transfer characteristics are presented in Fig. 3. More particularly, the LFN for the A300 GFET with $L=300\ nm$ is shown in Fig. 6a and 6d, the LFN for the A200 GFET with $L=200\ nm$ is shown in Fig. 6b and 6e and the LFN for the A100 GFET with $L=100\ nm$ is shown in Fig. 6c and 6f. Upper plots correspond to the LFN results from the highest and lowest drain voltage while lower plots to the rest of the drain voltages available. LFN data is represented by round square while solid lines is shown with solid lines. For comparison, long channel model proposed in ref. 35 is also shown in dashed lines. In fact, this long channel model equals to the sum of $\Delta NA$, $\Delta \mu A$ and $\Delta R$ terms of eqns (5,7, and 10), respectively. In the upper plots, it is clear that the proposed model coincides with the long channel model at low electric field regime where the VS effect is not significant. The agreement with the data is consistent apart from some regions away from CNP at n-type regime where the IV model was also not consistent (See Fig. 3) due to asymmetries of the data. At higher drain voltage region, the proposed short channel model is very accurate. It predicts a reduction of LFN in comparison with long channel model and this fully agrees with experimental data especially at CNP where $\Delta R$ is not significant. This phenomenon is analyzed and modeled for the first time. In the down plots of Fig. 6, the model is tested for the rest of drain voltages and it can be observed that its behavior remains precise. For drain voltages up to $0.1\ V$ the difference between the short and long channel model is negligible but above this value, VS effect significantly reduces LFN PSD and this is accurately captured by the proposed model. $\Delta R$ contribution is significant away from CNP especially for higher drain voltage levels where LFN PSD is almost constant and $\Delta R$ model successfully predicts this behavior. The IV and LFN models' validation for the remaining devices B300, B200 and B100 is presented in ESI F and G (Fig. S3 and S4 respectively).

The three extracted LFN parameters, $N_T$, $\alpha_H$ and $S_{\Delta R}^2$ are shown in Table 1. The value of $N_T$ ranges from ~$5.5 \cdot 10^{18}$ -$8 \cdot 10^{19}\ eV^{-1}cm^{-3}$ depending on the device and is a little lower than values extracted in other works[33-35] related to GFETs while also closer to typical values of MOSFETs[21-22, 42]. Regarding $\alpha_H$, is between ~$2.5 \cdot 10^{-4}$ - $5 \cdot 10^{-3}$ which is similar to ref. 35 and higher than MOSFETs[21-22, 42]. Finally, $S_{\Delta R}^2$ parameter ranges from ~$3.4 \cdot 10^{-3}$ -$2 \cdot 10^{-2}\ \Omega^2/Hz$ which is much more than MOSFETs[21, 42] but this is reasonable since the contact resistance $R_c$ of the measured GFETs is more intense than Si devices.

## Conclusions

In conclusion, a comprehensive physics-based analytical LFN model for short channel SL GFETs is proposed in the present study which is proved to be very consistent. This model extends the model derived in ref. 35 in order to deal with effects that are important when gate is scaled down, i.e. in high frequency devices. In particular, VS can contribute strongly to LFN at higher electric fields and short gate lengths. The analysis of VS effect and the way it affects drain current and consequently LFN, is thoroughly examined. Carrier number and mobility fluctuation mechanisms are the main contributors to LFN and VS phenomenon affects both of them. At low drain voltage regime, VS effect contribution is negligible since longitudinal electric field is much lower than critical field but for higher drain voltages, VS effect becomes significant and as a result, leads to a reduction of LFN especially at CNP. The proposed model is validated with novel experimental data from CVD grown GFETs[46-49] at three different short channel lengths ($L=300, 200, 100\ nm$) and in every case it accurately captures the reduction of LFN at higher drain voltages, something presented for the first time. At lower drain voltages, it coincides with the long channel model proposed in ref. 35. $\Delta N$ effect models the M-shape of the LFN data near CNP while $\Delta \mu$ effect contributes mainly near CNP. Moreover, a compact model for the effect of contact resistance on LFN is also derived with very consistent results at higher gate voltage regimes where contact resistance is dominant for GFETs. All the equations are solved in a compact way which makes the resulting model suitable for circuit simulators. In general, the extension of the well-established model of ref. 35 in order to cover the VS effect contribution to LFN is of high importance. Graphene is used in RF applications where the demand for high maximum oscillation frequencies makes essential the use of short channel devices and high drain voltage levels. The latter are the device dimensions and operating conditions where VS effect becomes important. The VS effect is found to reduce significantly the LFN.

## Experimental Data

### Devices fabrication

The main features of our GFETs structure are the bottom gates with native oxide. We designed different gate length of *100 nm*, *200 nm* and *300 nm*. The channel width is *2x12 µm*. CVD graphene grown on Cu foil[49] was used for wafer scale fabrication and good electrical properties. The double bottom-gates were patterned by using electron beam lithography (EBL), followed with *40 nm Al* deposition and lift-off process. After, the dielectric of $Al_2O_3$ (~ *4 nm* in thickness) was obtained by exposing the sample with bottom-gate structure to the air at room temperature. In this work, the bottom-gate structure was used in order to ease the natural oxidation process and avoid e-beam exposure on graphene channel. Monolayer graphene was transferred on top of the pre-patterned bottom-gates. Reactive ion etching (RIE) $O_2$ plasma was used to define the channel. Source and drain were obtained by depositing $Ni/Au$ (*20 nm/30 nm*) followed by a lift-off process. In order to make our devices compatible with on-chip probe measurements, the device fabrication was embedded in a *50 Ohm* coplanar waveguide (Fig. 1b). The waveguide is realized by EBL, followed by deposition of $Ni/Au$ (*50 nm/300 nm*).

### Electrical characterization

At each polarization, the drain-to-source current signal is measured with a custom-made current-to-voltage converter with two parallel inputs for DC (low-pass filter at *0. 1Hz* for I-V characteristics) and AC (band-pass filter from *0.1 Hz* to *7 kHz* for noise characterization). The data acquisition is performed using a National Instruments DAQ-card system (NI 6363). In order to stabilize the $I_{DS}$ current value at each gate bias, the sampling condition is $dI_{DS}/dt <1·10^7 A/s$ before each recorded point. For the noise characterization, the sampling frequency was set to *50 kHz* for a period of time of *13* seconds choosing the Welch's method in which 10 segments overlap by *50%*.

## Conflicts of interest

There are no conflicts to declare

## Acknowledgements

This work was funded by the Ministerio de Economía y Competitividad under the project TEC2015-67462-C2-1-R and the European Union's Horizon 2020 research and innovation program under Grant Agreement No. GrapheneCore2 785219 (Graphene Flagship), Marie Skłodowska-Curie Grant Agreement No 665919 and Grant Agreement No. 732032 (BrainCom). This work was also partly supported by the French RENATECH network. The ICN2 is supported by the Severo Ochoa Centres of Excellence programme, funded by the Spanish Research Agency (AEI, grant no. SEV-2017-0706).

## References

(1)  K. S. Novoselov, A. K. Geim, S. V. Morozov, Y. Zhang, S. V. Dubonos, I. V. Grigorieva and A. A. Firsov, *Science*, 2004, **306**, 666.

(2)  A. K. Geim and K. S. Novoselov, *Nature Materials*, 2007, **6**, 183-191.

(3)  F. Schwierz, *Nature Nanotechnology*, 2010, **5**, 487-496.

(4)  F. Schwierz, R. Granzner and J. Pezoldt, *Nanoscale*, 2015, **7**, 1567-1584.

(5)  J. S. Moon, D. Curtis, D. Zehnder, S. Kim, D. K. Gaskill, G. G. Jernigan, R. L. Myers-Ward, C. R. Eddy, P. M. Campbell, K.-M. Lee and P. Asbeck, *IEEE Electron Device Letters*, 2011, **2(3)**, 270-272.

(6)  E. Kougianos, S. Joshi and S. P. Mohanty, *IEEE 7th Computer Society Annual Symposium on VLSI (ISVLSI)*, 2015.

(7)  L. Vicarelli, M. S. Vitiello, D. Coquillat, A. Lombardo, A. C. Ferrari, W. Knap, M. Polini, V. Pellegrini and A. Tredicucci, *Nature Mater.*, 2012, **11**, 865-871.

(8)  G. Auton, D. B. But, J. Zhang, E. Hill, D. Coquillat, C. Consejo, P. Nouvel, W. Knap, L. Varani, F. Teppe, J. Torres and A. Song, *Nano Letters*, 2017, **17(11)**, 7015-7020.

(9)  X. Yang, A. Vorobiev, K. Jeppson, J, Stake, L. Banszerus, C. Stampfer, M. Otto and D. Neumaier, *IEEE 43rd International Conference on Infrared, Millimeter, and Terahertz Waves (IRMMW-THz)*, 2018.

(10) F. Schedin, A. K. Geim, S. V. Mozorov, E. W. Hill, M. I. Blake, M. I. Katsnelson and K. S. Novoselov, *Nature Mater.*, 2007, **16**, 652-655.

(11) L. Hess, M. Seifert and J. A. Garrido, *Proc. IEEE*, 2013, **101(7)**, 1780-1792.

(12) S. Rumyantsev, G. Liu, M. S. Shur, R. A. Potyrailo and A. A. Balandin, *Nano Letters*, 2012, **12(5)**, 2294-2298.

(13) C. Herbert, E. Masvidal-Codina, A. Suarez-Perez, A. Bonaccini-Calia, G. Piret, R. Garcia-Cortadella, X. Illa, Del E. Corro Garcia, J. M. De la Cruz Sanchez, D. V. Casals, E. Prats-Alfonso, J. Bousquet, P. Godignon, B. Yvert, R. Villa, M. V. Sanchez-Vives, A. Guimera-Brunet and J. A. Garrido, *Adv. Funct. Mater.*, 2017, **1703976**.

(14) F. Bonaccorso, Z. Sun, T. Hasan and A. C. Ferrari, *Nature Photonics*, 2010, **4**, 611-622.

(15) S. K. Lee, C. G. Kang, Y. G. Lee, C. Cho, E. Park, H. J. Chung, S. Seo, H. D. Lee and B. H. Lee, *Carbon*, 2012, **50(11)**, 4046-4051.

(16) A. L. McWhorter, *Semiconductor Surface Physics*, 1957, 207-228.

(17) M. J. Uren, D. J. Day and M. J. Kirton, *Applied Physics Letters*, 1985, **47(11)**, 1195-1197.

(18) F. N. Hooge, *Physica*, 1976, **83B**, 14-23.

(19) G. Reimbold, *IEEE Transactions on Electron Devices*, 1984, **31(9)**, 1190-1194.

(20) G. Ghibaudo, *Solid State Electronics*, 1989, **32(7)**, 563-565.

(21) K. K. Hung, P. K. Ko, C. Hu and Y. C. Cheng, *IEEE Transactions on Electron Devices*, 1990, **37(3)**, 654-665.

(22) N. Mavredakis, A. Antonopoulos and M. Bucher, *5th IEEE Europ. Conf. on Circuits & Systems for Communications (ECCSC)*, 2010, 86-89.

(23) A. A. Balandin, *Nature Nanotechnology*, 2013, **8**, 549-555.

(24) G. Liu, S. Rumyantsev, M. S. Shur and A. A. Balandin, *Applied Physics Letters*, 2013, **102(9)**, 93111.

(25) G. Xu, C. M. Torres, Y. Zhang, F. Liu, E. B. Song, M. Wang, Y. Zhou, C. Zeng and K. L. Wang, *Nano Letters*, 2010, **10(9)**, 3312-3317.

(26) Y. Zhang, E. E. Mendez and X. Du, *ACS Nano*, 2011, **5(10)**, 8124-8130.

(27) A. N. Pal, S. Ghatak, V. Kochat, A. Sneha, A. Sampathkumar, S. Raghavan and A. Ghosh, *ACS Nano*, 2011, **5(3)**, 2075-2081.

(28) S. Rumyantsev, G. Liu, W. Stillman, M. Shur and A. A. Balandin, *Journal Phys. Cond. Matter*, 2010, **22(39)**, 395302(1-7).

(29) N. Sun, K. Tahy, H. Xing, D. Jena, G. Arnold and S. T. Ruggiero, *J. Low Temp. Phys.*, 2013, **172(3-4)**, 202-211.

(30) B. Pellegrini, *Eur. Phys. J. B.*, 2013, **86** : 373.

(31) B. Pellegrini, P. Marconcini, M. Macucci, G. Fiori and G. Basso, *Journal of Statistical Mechanics: Theory and Experiment*, 2016, 054017.

(32) C. Mukherjee, J. D. Aguirre-Morales, S. Fregonese, T. Zimmer, C. Maneux, H. Happy and W. Wei, *45th IEEE Europ. Solid State Dev. Res. Conf. (ESSDERC)*, 2015, 176-179.

(33) S. Peng, Z. Jin, D. Zhang, J. Shi, D. Mao, S. Wang and G. Yu, *ACS Appl. Mater. Interfaces*, 2017, **9(8)**, 6661-6665.

(34) I. Heller, S. Chatoor, J.Mannik, M. A. G. Zevenbergen, J. B. Oostinga, A. F. Morpurgo, C. Dekker, and S. G. Lemay, *Nano Letters*, 2010, **10(5)**, 1563-1567.

(35) N. Mavredakis, R. Garcia Cortadella, A. Bonaccini Calia, J. A. Garrido and D. Jimenez, *Nanoscale*, 2018, **10**, 14947-14956.

(36) A. S. Roy, Thesis N0 3921, 2007, EPFL.

(37) I. Meric, M. Y. Han, A. F. Young, B. Ozyilmaz, P. Kim and K. L. Shepard, *Nature Nanotechnology*, 2008, 3, 654-659.

(38) J. Chauhan and J. Guo, *Applied Physics Letters*, 2009, **95(2)**, 023120.

(39) V. E. Dorgan, M. H. Bae and E. Pop, *Applied Physics Letters*, 2010, **97(8)**, 082112.

(40) S. Thiele and F. Schwierz, *Journal of Applied Physics*, 2011, **110(3)**, 034506.

(41) A. Barreiro, M. Lazzeri, J. Moser, F. Mauri and A. Bachtold, *Phys. Rev. Lett.*, 2009, **103**, 076601.

(42) C. Enz and E. Vitoz, *John Wiley and Sons*, 2006.

(43) D. Jiménez and O. Moldovan, *IEEE Transactions on Electron Devices*, 2011, **58(11)**, 4377-4383.

(44) G. M. Landauer, D. Jiménez, and J. L. Gonzalez, *IEEE Trans. Nanotechnolog.*, 2014, **13(5)**, 895-904.

(45) F. Pasadas and D. Jiménez, *IEEE Transactions on Electron Devices*, 2016, **61(7)**, 2936-2941.

(46) W. Wei, X. Zhou, G. Deokar, H. Kim, M. M. Belhaj, E. Galopin, E. Pallecchi, D. Vignaud and H. Happy, *IEEE Transactions on Electron Devices*, 2011, **62(9)**, 2769-2773.

(47) W. Wei, E. Pallecchi, S. Hague, S. Borini, V. Avramovic, A. Centeno, Z. Amaia and H. Happy, *Nanoscale*, 2016, **8**, 14097-14103.

(48) W. Wei, D. Fadil, E. Pallecchi, G. Dambrine, H. Happy, M. Deng, S. Fregonese and T. Zimmer, *24th IEEE Int. Conf. on Noise and Fluctuations (ICNF)*, 2017.

(49) G. Deokar, J. Avila, I. Razado-Colambo, J. L. Cordon, C. Boyaval, E. Galopin, M. C. Asencio and D. Vignaud, *Carbon*, 2015, **89**, 82.

(50) F. Chaves, D. Jimenez, A. A. Sagade, W. Kim, J. Riikonen, H. Lipsanen and D. Neumaier, *2D Materials*, 2015, **2(2)**, 025006.

# Supplementary Information for:

# Velocity Saturation effect on Low Frequency Noise in short channel Single Layer Graphene FETs

Nikolaos Mavredakis*[a], Wei Wei[b], Emiliano Pallecchi[b], Dominique Vignaud[b], Henri Happy[b], Ramon Garcia Cortadella[c], Andrea Bonaccini Calia[c], Jose A. Garrido[c, d] and David Jiménez[a]

[a] Departament d'Enginyeria Electrònica, Escola d'Enginyeria, Universitat Autònoma de Barcelona, Bellaterra 08193, Spain

[b] Institute of electronics, Microelectronics and Nanotechnology, CNRS UMR8520, 59652 Villeneuve d'Ascq, France.

[c] Catalan Institute of Nanoscience and Nanotechnology (ICN2), CSIC, Barcelona Institute of Science and Technology, Campus UAB, Bellaterra, Barcelona, Spain

[d] ICREA, Pg. Lluis Companys 23, 08010 Barcelona, Spain

* nikolaos.mavredakis@uab.es

## A. Supplementary Information: Drift-Diffusion current equation when Velocity Saturation effect is included. How the integral variable changes from $dx$ to $dV_c$.

Eqn (1) of the main manuscript calculates the drain current $I_D$ of the device when Velocity Saturation (VS) effect is taken into consideration. It is known that[43-45]:

$$E = -\frac{dV}{dx}, \quad E_x = -\frac{d\psi}{dx} = -\frac{dV + dV_c}{dx}, \quad \frac{dV}{dV_c} = \frac{C_q + C}{C}, \quad E_c = \frac{u_{sat}}{\mu} \tag{Eq.A1}$$

where all the above quantities are defined in the main manuscript apart from $\psi$ which is the electrostatic potential. If eqn (A1) is replaced into the effective mobility term of eqn (1) of the main manuscript and if this is then inserted into drain current term of eqn (1) of the main manuscript then the drain current $I_D$ is given as:

$$I_D = -WQ_{gr}\frac{\mu}{1+\frac{\mu}{\upsilon_{sat}}\frac{\left|-dV-dV_c\right|}{dx}}\frac{dV}{dx} = -WQ_{gr}\frac{\mu}{1+\frac{\mu}{\upsilon_{sat}}\frac{\left|-dV\left(1+\frac{dV_c}{dV}\right)\right|}{dx}}\frac{dV}{dx} = -WQ_{gr}\frac{\mu}{1+\frac{\mu}{\upsilon_{sat}}\frac{\left|-dV\right|}{dx}\left(\frac{C_q+2C}{C_q+C}\right)}\frac{dV}{dx} \tag{Eq.A2}$$

Then from eqns (A1, A2) we can end up with eqn (2) of the main manuscript with $k=2 \cdot e^3/(\pi \cdot h^2 \cdot u^2 f)$ [43-44] where $u_f$ is the Fermi velocity (=$10^6$ m/s), $h$ the reduced Planck constant (=$1,05 \cdot 10^{-34}$ J·s). Bias dependent term $g(V_c)$ which expresses the normalized drain current $I_D$ is calculated as[43]:

$$gV_c = \left[g\left(V_c\right)\right]_{V_{cs}}^{V_{cd}} + \frac{\alpha V_{DS}}{k} = \frac{V_{cs}^3 - V_{cd}^3}{3} + \frac{k}{4C}\left[\text{sgn}\left(V_{cd}\right)V_{cd}^4 - \text{sgn}\left(V_{cs}\right)V_{cs}^4\right] + \frac{\alpha V_{DS}}{k}, \quad I_D = \frac{\mu Wk}{2L_{eff}}gV_c \tag{Eq.A3}$$

while graphene charge is given by[43-45]:

$$Q_{gr} = \frac{k}{2}\left(V_c^2 + \alpha / k\right) \tag{Eq.A4}$$

and chemical potential at source and drain as[43]:

$$V_{cs,d} = \frac{C - \sqrt{C^2 \pm 2k\left[C_{top}\left(V_{Gtop} - V_{Gtop0} - V_{s,d}\right) + C_{back}\left(V_{Gback} - V_{Gback0} - V_{s,d}\right)\right]}}{\pm k} \tag{Eq.A5}$$

where $\alpha = 2.\rho_0.e$ is a residual charge ($\rho_0$) related term, $V_{Gtop}$-$V_{Gtop0}$ and $V_{Gback}$-$V_{Gback0}$ are the top- and back-gate voltage overdrives, respectively. The discrimination between positive and negative $V_{DS}$ defines the sign of electrical field $E$ and consequently the signs of second terms of the right hand of eqn (2) of the main manuscript. In more detail, if $V_{DS}$>0 then $dV$<0 and $E$=-$dV/dx$>0 (top branch of eqn (2)) while if $V_{DS}$<0 then $dV$>0 and $E$=-$dV/dx$<0 (bottom branch of eqn (2)). This relation between $dx$ and $dV_c$ is very crucial for the calculations of LFN as it will be shown later[35].

**B. Supplementary Information: $L_{eff}$ calculation**

In the denominator of eqn (3) of the main manuscript, $L_{eff}$ is defined which represents an effective length to take into account VS effect. The thorough procedure of its extraction will be presented below. From eqn (A2) we can take:

$$I_D + \frac{I_D \mu}{\upsilon_{sat}}\frac{|-dV|}{dx}\left(\frac{C_q + 2C}{C_q + C}\right) = -WQ_{gr}\mu\frac{dV}{dx} \Leftrightarrow I_D dx + \frac{I_D \mu}{\upsilon_{sat}}|-dV|\left(\frac{C_q + 2C}{C_q + C}\right) = -WQ_{gr}\mu dV \tag{Eq.A6}$$

If we integrate each term of eqn (A6) from S (Source) to D (Drain):

$$I_D L + \int_{V_s}^{V_d}\frac{I_D \mu}{\upsilon_{sat}}\left(\frac{C_q + 2C}{C_q + C}\right)\Big||-dV| = -\int_{V_s}^{V_d} WQ_{gr}\mu dV \tag{Eq.A7}$$

Again the sign of $V_{DS}$ defines the sign of the second left hand term in eqn (A7). If $V_{DS}$>0 then $dV$<0 and -$dV$>0 and if eqn (A1) is also taken into account, eqn (A7) becomes:

$$I_D = \frac{-\int_{V_s}^{V_d} WQ_{gr}\mu dV}{L - \int_{V_s}^{V_d}\frac{\mu}{\upsilon_{sat}}\left(\frac{C_q + 2C}{C_q + C}\right)dV} = \frac{W\mu\int_{V_{cd}}^{V_{cs}} Q_{gr}\frac{C_q + C}{C}dV_c}{L + \int_{V_{cd}}^{V_{cs}}\frac{\mu}{\upsilon_{sat}}\left(\frac{C_q + 2C}{C}\right)dV_c} \tag{Eq.A8}$$

On the other hand, if $V_{DS}$<0 then $dV$>0 and -$dV$<0 and similarly to before eqn (A7) becomes:

$$I_D = \frac{-\int_{V_s}^{V_d} WQ_{gr}\mu dV}{L + \int_{V_s}^{V_d}\frac{\mu}{\upsilon_{sat}}\left(\frac{C_q + 2C}{C_q + C}\right)dV} = \frac{W\mu\int_{V_{cd}}^{V_{cs}} Q_{gr}\frac{C_q + C}{C}dV_c}{L + \int_{V_{cs}}^{V_{cd}}\frac{\mu}{\upsilon_{sat}}\left(\frac{C_q + 2C}{C}\right)dV_c} \tag{Eq.A9}$$

Eqn (A8, A9) are identical to eqn (3) of the main manuscript. It is very significant to solve the denominator integrals of the above equations analytically in order to use them in LFN expressions where $L_{eff}$ is present. The positive drain voltage case will be shown while the negative one can be solved similarly. To proceed with the calculation, two cases should be discriminated regarding $u_{sat}$ value as described in eqn (4) of the main manuscript; one for $V_c < V_{ccrit}$ where $u_{sat}$ is constant and the other for the opposite conditions where $u_{sat}$ is inversely proportional to $sqrt(V_c^2 + a/k)$. For the first case where $u_{sat}$ is constant we take:

$$L_{eff} = L + \frac{\mu}{SC}\left[2CV_c \pm \frac{1}{2}kV_c^2\right]_{V_{cd}}^{V_{cs}}$$  (Eq.A10)

while for the second case where $u_{sat}$ is inversely proportional to $sqrt(V_c^2 + a/k)$ we have:

$$L_{eff} = L + \frac{\mu}{NC}\left[\pm\frac{1}{3}\sqrt{V_c^2 + \frac{\alpha}{k}}\left(\alpha + V_c\left(\pm 3C + kV_c\right)\right) + \frac{\alpha C \ln\left(V_c + \sqrt{V_c^2 + \frac{\alpha}{k}}\right)}{k}\right]_{V_{cd}}^{V_{cs}}$$  (Eq.A11)

where $S$, $N$ are defined in eqn (4) of the main manuscript.

* $\pm, \mp$ : Top sign refers to $V_c > 0$ and bottom sign to $V_c < 0$.

## C. Supplementary Information: Detailed examination of graphene chemical potential and longitudinal electric field locally in the channel.

The behavior of significant quantities such as chemical potential or longitudinal electric field along the channel at different bias conditions regarding gate and drain voltage, can contribute to the understanding of the operation of the device and clarify the way that short channel effects such as VS can influence this operation. In Fig. S1, the chemical potential $V_c(x)$ and the longitudinal electric field $E_x(x)$ are presented vs. channel position $x$ at charge neutrality point – CNP (a,d), away from CNP in p-type region (b,e) and away from CNP in n-type region (c, f) for a channel length of $L=100$ $nm$ for both low and high drain voltage values ($V_{DS}$=60 mV, 0.3 V). Regarding $V_c(x)$, it can be easily observed that it is almost identical at every position at low drain voltage which indicates a uniform channel while as the drain voltage gets higher this homogeneity is not valid anymore since the fluctuation of $V_c(x)$ is quite large. As far as $E_x(x)$ is concerned, it is much lower than critical field $E_c(x)$ at low drain voltage and this is the reason why VS effect is negligible there while at high drain voltage $E_x(x)$ becomes comparable to $E_c(x)$ and this affects the operation of the device due to the degradation of effective mobility as it is illustrated in Fig. 2c of the main manuscript. Moreover, *LFN* is also affected as it will be shown in the next section.

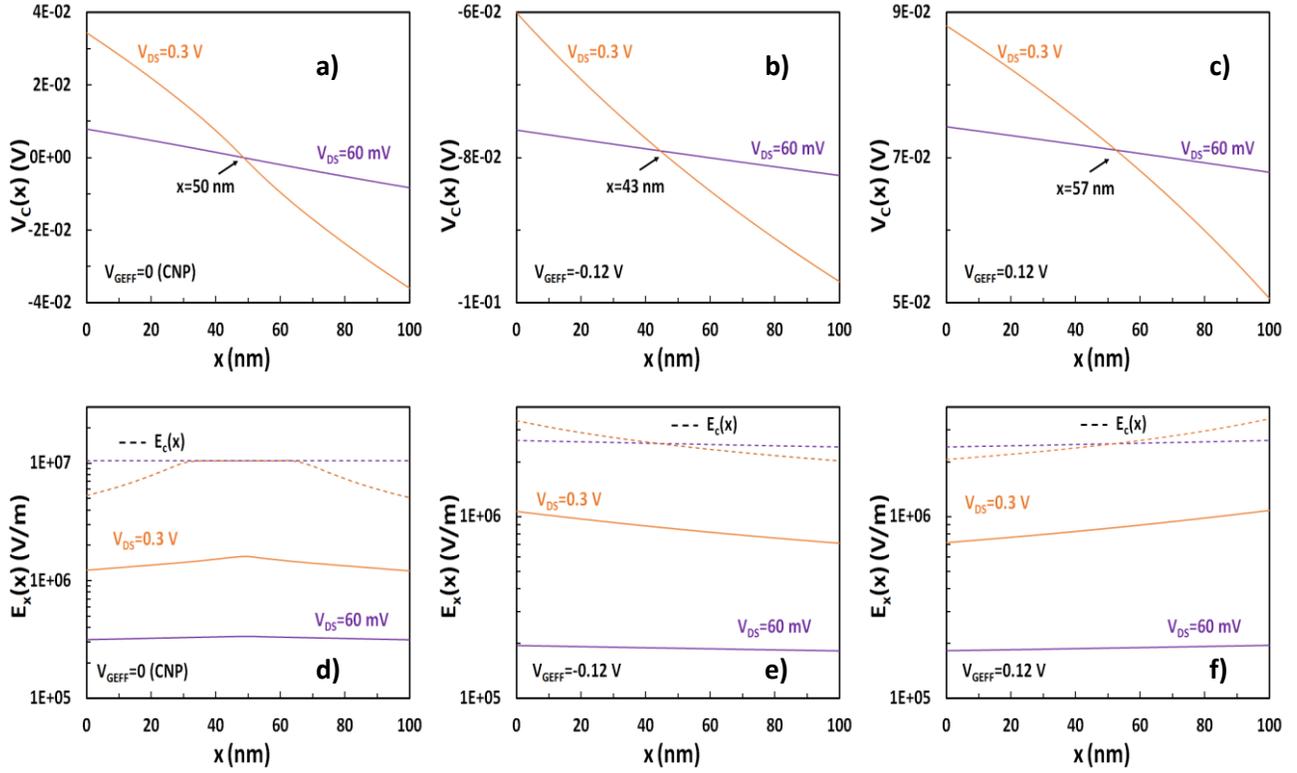

**Fig. S1** Graphene chemical potential $V_c(x)$ (upper plots) and longitudinal electric field $E_x(x)$ (down plots) vs. channel position $x$, for a, d) $V_{GEFF}=0\ V\ (CNP)$, b, e) $V_{GEFF}=-0,12\ V$ and c,f) $V_{GEFF}=0,12\ V$ at $V_{DS}=60\ mV$, $0.3\ V$ for $W/L=12\ \mu m/100\ nm$.

**D. Supplementary Information: Thorough procedure for calculation of local *LFN* and examination of VS effect on it for both carrier number and mobility fluctuations effect.**

As described in ref. 35, the *LFN* methodology applied here considers a noiseless channel apart from an elementary slice between $x$ and $x+\Delta x$. This local noise contribution can be represented by a local current noise source with a *PSD* $S_{\delta I^2 n}$. Without entering into much detail since these are described thoroughly in ref. 35, the *PSD* of the total noise current fluctuation at the drain side $S_{ID}$ due to all different sections along the channel is obtained by summing their elementary contributions $S_{\delta I^2_{nD}}$ assuming that the contribution of each slice at different positions along the channel remains uncorrelated[35, 42]:

$$S_{ID} = \int_0^L \frac{S_{\delta I^2_{nD}}(\omega, x)}{\Delta x}dx = \int_0^L G_{CH}^2 \Delta R^2 \frac{S_{\delta I^2_n}(\omega, x)}{\Delta x}dx = \frac{1}{L_{eff}^2}\int_0^L \Delta x S_{\delta I^2_n}(\omega, x)\,dx \qquad (Eq.A12)$$

since with VS effect included we obtain[36, 42]:

$$G_{CH}^2 \Delta R^2 = \left(\left(\frac{dI_d}{dV}\right)^2\left(\frac{\Delta V}{I_d}\right)^2\right) = \left(\frac{W\mu_{eff}Q_{gr}}{L_{eff}}\frac{\Delta x}{W\mu_{eff}Q_{gr}}\right)^2 = \left(\frac{\Delta x}{L_{eff}}\right)^2 \qquad (Eq.A13)$$

and

$$S_{\delta I_{sD}^2}\left(\omega,x\right)=G_{CH}^2\Delta R^2 S_{\delta I_n^2}\left(\omega,x\right) \tag{Eq.A14}$$

Regarding carrier number fluctuation effect, the *PSD* of the normalized local noise source divided by squared drain current is given in ref. 35 (eqn (4)-pp 10). If this is inserted in eqn (A12) above then the integral below expresses the total *LFN PSD* normalized with squared drain current at *1 Hz* regarding *ΔN* mechanism:

$$\left.\frac{S_{I_D}}{I_D^2}f\right|_{\Delta N}=\frac{KT\lambda N_T}{WL_{eff}^2}\int_0^L\left(\frac{e}{Q_{gr}}\frac{C_q}{C_q+C}\right)^2 dx \tag{Eq.A15}$$

As the *LFN* model is based on a chemical potential based model[43-45], the integral variable should change from *x* to $V_c$ according to eqn (2) of the main manuscript and this is the point where VS effect enters noise calculations. We will proceed with the case of $V_{DS}>0$ but the procedure is similar for a negative $V_{DS}$. After applying eqn (2) of the main manuscript at eqn (A15):

$$\left.\frac{S_{I_D}}{I_D^2}f\right|_{\Delta N}=\frac{KT\lambda N_T}{WL_{eff}^2}\int_{V_{cd}}^{V_{cs}}\left[\left(\frac{e}{Q_{gr}}\right)^2\left(\frac{C_q}{C_q+C}\right)^2\right]\left[\frac{Q_{gr}\,2L_{eff}}{kg_{vc}}\left(\frac{C_q+C}{C}\right)-\frac{\mu}{\upsilon_{sat}}\left(\frac{C_q+2C}{C}\right)\right]dV_c \tag{Eq.A16}$$

Eqn (A16) can be split into two integrals as it is shown below:

$$\left.\frac{S_{I_D}}{I_D^2}f\right|_{\Delta NA}=\frac{KT\lambda N_T}{WL_{eff}^2}\int_{V_{cd}}^{V_{cs}}\left(\frac{e}{Q_{gr}}\right)^2\left(\frac{C_q}{C_q+C}\right)^2\frac{Q_{gr}\,2L_{eff}}{kg_{vc}}\left(\frac{C_q+C}{C}\right)dV_c \tag{Eq.A17}$$

and

$$\left.\frac{S_{I_D}}{I_D^2}f\right|_{\Delta NB}=\frac{KT\lambda N_T}{WL_{eff}^2}\int_{V_{cd}}^{V_{cs}}\left(\frac{e}{Q_{gr}}\right)^2\left(\frac{C_q}{C_q+C}\right)^2\frac{\mu}{\upsilon_{sat}}\left(\frac{C_q+2C}{C}\right)dV_c \tag{Eq.A18}$$

If $Q_{gr}$ is replaced by eqn (A4) and $C_q=k|V_c|$[35, 43] then we have:

$$\left.\frac{S_{I_D}}{I_D^2}f\right|_{\Delta NA}=\frac{A1}{g_{vc}}\int_{V_{cd}}^{V_{cs}}\frac{V_c^2}{\left(V_c^2+\alpha/k\right)\left(C+k|V_c|\right)}dV_c \quad with \quad A1=\frac{4KT\lambda N_T e^2}{CWL_{eff}} \tag{Eq.A19}$$

and

$$\left.\frac{S_{I_D}}{I_D^2}f\right|_{\Delta NB}=B1\int_{V_{cd}}^{V_{cs}}\frac{V_c^2}{u_{sat}\left(V_c^2+\alpha/k\right)^2}\frac{k|V_c|+2C}{\left(k|V_c|+C\right)^2}dV_c \quad with \quad B1=\frac{A1\mu}{L_{eff}} \tag{Eq.A20} \text{ Total }\Delta N$$

*LFN* at *1 Hz* is then given:

$$\left.\frac{S_{I_D}}{I_D^2}f\right|_{\Delta N}=\left.\frac{S_{I_D}}{I_D^2}f\right|_{\Delta NA}-\left.\frac{S_{I_D}}{I_D^2}f\right|_{\Delta NB} \tag{Eq.A21}$$

As far as mobility fluctuation effect is concerned, a similar process is followed where:

$$\left.\frac{S_{I_D}}{I_D^2}f\right|_{\Delta\mu} = \frac{\alpha_H e}{WL_{eff}^2}\int_0^L\frac{1}{Q_{gr}}dx \tag{Eq.A22}$$

The integral variable changes from $x$ to $V_c$ according to eqn (2) of the main manuscript and eqn (A22) becomes:

$$\left.\frac{S_{I_D}}{I_D^2}f\right|_{\Delta\mu} = \frac{\alpha_H e}{WL_{eff}^2}\int_{V_{cd}}^{V_{cs}}\frac{1}{Q_{gr}}\left[\frac{Q_{gr}\,2L_{eff}}{kg_{vc}}\left(\frac{C_q+C}{C}\right)-\frac{\mu}{\upsilon_{sat}}\left(\frac{C_q+2C}{C}\right)\right]dV_c \tag{Eq.A23}$$

Similarly as before, eqn (A23) is split into two new integrals:

$$\left.\frac{S_{I_D}}{I_D^2}f\right|_{\Delta\mu A} = \frac{\alpha_H e}{WL_{eff}^2}\int_{V_{cd}}^{V_{cs}}\frac{1}{Q_{gr}}\left[\frac{Q_{gr}\,2L_{eff}}{kg_{vc}}\left(\frac{C_q+C}{C}\right)\right]dV_c \tag{Eq.A24}$$

and

$$\left.\frac{S_{I_D}}{I_D^2}f\right|_{\Delta\mu B} = \frac{\alpha_H e}{WL_{eff}^2}\int_{V_{cd}}^{V_{cs}}\frac{1}{Q_{gr}}\frac{\mu}{\upsilon_{sat}}\left(\frac{C_q+2C}{C}\right)dV_c \tag{Eq.A25}$$

again If $Q_{gr}$ is replaced by eqn (A4) and $C_q=k/|V_c|^{35,\,43}$ then we have:

$$\left.\frac{S_{I_D}}{I_D^2}f\right|_{\Delta\mu A} = \frac{A2}{g_{vc}}\int_{V_{cd}}^{V_{cs}}\left(C+k|V_c|\right)dV_c \quad with \quad A2=\frac{2\alpha_H e}{WL_{eff}Ck} \tag{Eq.A26}$$

and

$$\left.\frac{S_{I_D}}{I_D^2}f\right|_{\Delta\mu B} = B2\int_{V_{cd}}^{V_{cs}}\frac{k|V_c|+2C}{u_{sat}\left(V_c^2+a/k\right)}dV_c \quad with \quad B2=\frac{A2\mu}{L_{eff}} \tag{Eq.A27}$$

Total $\Delta\mu$ LFN at $1\ Hz$ is given:

$$\left.\frac{S_{I_D}}{I_D^2}f\right|_{\Delta\mu} = \left.\frac{S_{I_D}}{I_D^2}f\right|_{\Delta\mu A} - \left.\frac{S_{I_D}}{I_D^2}f\right|_{\Delta\mu B} \tag{Eq.A28}$$

Terms $\Delta NB$, $\Delta\mu B$ change depending on the above condition. For the first case where $u_{sat}$ is constant we take:

$$\left.\frac{S_{I_D}}{I_D^2}f\right|_{\Delta NB} = \frac{B1}{S}\int_{V_{cd}}^{V_{cs}}\frac{V_c^2}{\left(V_c^2+\alpha/k\right)^2}\frac{k|V_c|+2C}{\left(k|V_c|+C\right)^2}dV_c \tag{Eq.A29}$$

and

$$\left.\frac{S_{I_D}}{I_D^2}f\right|_{\Delta\mu B} = \frac{B2}{S}\int_{V_{cd}}^{V_{cs}}\frac{k|V_c|+2C}{V_c^2+\alpha/k}dV_c \tag{Eq.A30}$$

while for the second case where $u_{sat}$ is inversely proportional to $sqrt(V_c^2+a/k)$ we have:

$$\frac{S_{I_D}}{I_D^2} f\Big|_{\Delta NB} = \frac{B1}{N} \int_{V_{cd}}^{V_{cs}} \frac{V_c^2}{\left(V_c^2 + \alpha/k\right)^{3/2}} \frac{k|V_c| + 2C}{\left(k|V_c| + C\right)^2} dV_c \qquad \text{(Eq.A31)}$$

and

$$\frac{S_{I_D}}{I_D^2} f\Big|_{\Delta \mu B} = \frac{B2}{N} \int_{V_{cd}}^{V_{cs}} \frac{k|V_c| + 2C}{\sqrt{V_c^2 + \alpha/k}} dV_c \qquad \text{(Eq.A32)}$$

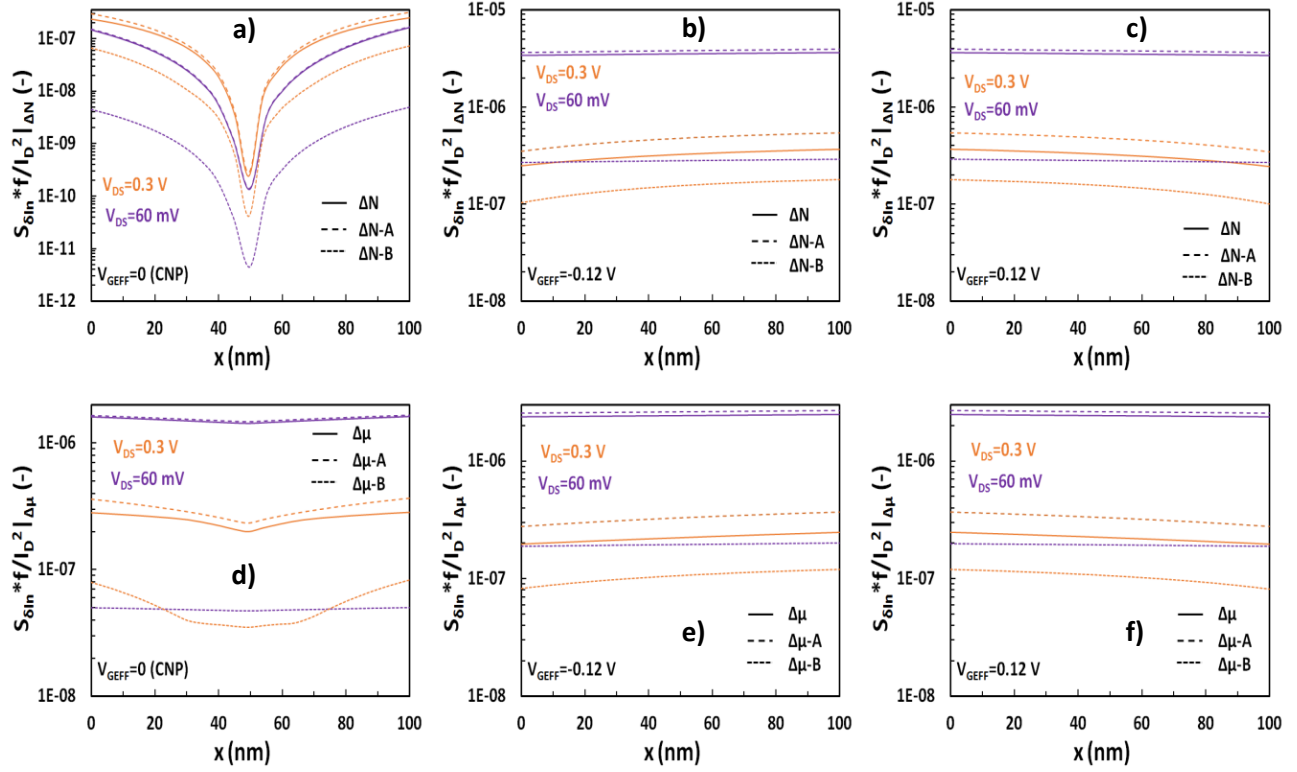

**Fig. S2** Normalized drain current noise divided by squared drain current referred to *1 Hz*, $S_{\delta in} f / I_D^2$, vs. channel potential *x* for *ΔN* (upper plots) and *Δμ* (down plots) noise mechanisms for a, d) $V_{GEFF}$=0 V (CNP), b, e) $V_{GEFF}$=-0,12 V and c,f) $V_{GEFF}$=0,12 V at $V_{DS}$=60 mV, 0.3 V for *W/L*=12 μm/100 nm. *ΔNA*, *ΔNB* contributors of *ΔN* effect are also shown in upper plots while *ΔμA*, *ΔμB* contributors of *Δμ* effect are also shown in down plots respectively.

For simplicity the terms of eqns (A19, A20, A21, A26, A27, A28) from now on will be referred as *ΔNA*, *ΔNB*, *ΔN*, *ΔμA*, *ΔμE*, *Δμ* respectively. In the case of a negative $V_{DS}$, the above integrals are transformed depending on eqn (2) of the main manuscript. It is important to mention here that terms *ΔNA*, *ΔμA* are identical to those extracted in ref. 35 where the long channel *LFN* model was presented. Terms *ΔNB*, *ΔμB* deal with the VS effect on *LFN* and this is clear since $u_{sat}$ term is included in eqns (A20, A27). As it was described before, $u_{sat}$ is constant for $V_c < V_{ccrit}$ and is inversely proportional to *sqrt($V_c^2 + a/k$)* otherwise.

What is inside all the above integrals expresses the different local noise sources which are shown vs. channel position $x$ in Fig. S2. In upper plots, $\Delta NA$, $\Delta NB$ and $\Delta N$ local noise terms are shown while in down plots $\Delta \mu A$, $\Delta \mu B$ and $\Delta \mu$ local noise terms are presented. Plots (a, d) correspond to charge neutrality point – CNP, plots (b,e) away from CNP in p-type region and plots (c, f) away from CNP in n-type region for a channel length of $L=100\ nm$ for both low and high drain voltage value ($V_{DS}=60\ mV,\ 0.3\ V$). A very important first observation is that $\Delta NB$, $\Delta \mu B$ terms related to the contribution of VS effect to $LFN$ are almost negligible at low $V_{DS}$ and there $\Delta NA \approx \Delta N$, $\Delta \mu A \approx \Delta \mu$ both near and away CNP. As $V_{DS}$ increases both $\Delta N$ and $\Delta \mu$ local noise values are decreased along the whole channel for every level of gate voltage. As it can be seen in the plots, orange solid lines become lower than orange dashed lines.

## E. Supplementary Information: Details on the integration procedure

To understand better the process that integrals of eqns (A19, A26 and A29-A32) are solved and lead to the analytical expressions of eqns (5-8) of the main manuscript, the following cases are taken:

$$0 \leq V_{cd} \leq V_{cs} \leq V_{ccrit} \rightarrow \int_{V_{cd}}^{V_{cs}} LFN = \int_{V_{cd}}^{V_{cs}} LFN_{Near\,CNP+},\ 0 \leq V_{ccrit} \leq V_{cd} \leq V_{cs} \rightarrow \int_{V_{cd}}^{V_{cs}} LFN = \int_{V_{cd}}^{V_{cs}} LFN_{Away\,CNP+}$$

$$0 \geq V_{cs} \geq V_{cd} \geq -V_{ccrit} \rightarrow \int_{V_{cd}}^{V_{cs}} LFN = \int_{V_{cd}}^{V_{cs}} LFN_{Near\,CNP-},\ 0 \geq -V_{ccrit} \geq V_{cs} \geq V_{cd} \rightarrow \int_{V_{cd}}^{V_{cs}} LFN = \int_{V_{cd}}^{V_{cs}} LFN_{Away\,CNP-}$$

$$0 \leq V_{cd} \leq V_{ccrit} \leq V_{cs} \int_{V_{cd}}^{V_{cs}} LFN = \int_{V_{cd}}^{V_{ccrit}} LFN_{Near\,CNP+} + \int_{V_{ccrit}}^{V_{cs}} LFN_{Away\,CNP+}$$

$$0 \geq V_{cs} \geq -V_{ccrit} \geq V_{cd} \rightarrow \int_{V_{cd}}^{V_{cs}} LFN = \int_{V_{cd}}^{-V_{ccrit}} LFN_{Away\,CNP-} + \int_{-V_{ccrit}}^{V_{cs}} LFN_{Near\,CNP-}$$

$$-V_{ccrit} \leq V_{cd} \leq 0 \leq V_{cs} \leq V_{ccrit} \rightarrow \int_{V_{cd}}^{V_{cs}} LFN = \int_{V_{cd}}^{0} LFN_{Near\,CNP-} + \int_{0}^{V_{cs}} LFN_{Near\,CNP+}$$

$$V_{cd} \leq -V_{ccrit} \leq 0 \leq V_{cs} \leq V_{ccrit} \rightarrow \int_{V_{cd}}^{V_{cs}} LFN = \int_{V_{cd}}^{-V_{ccrit}} LFN_{Away\,CNP-} + \int_{-V_{ccrit}}^{0} LFN_{Near\,CNP-} + \int_{0}^{V_{cs}} LFN_{Near\,CNP+}$$

$$V_{cd} \leq -V_{ccrit} \leq 0 \leq V_{ccrit} \leq V_{cs} \rightarrow \int_{V_{cd}}^{V_{cs}} LFN = \int_{V_{cd}}^{-V_{ccrit}} LFN_{Away\,CNP-} + \int_{-V_{ccrit}}^{0} LFN_{Near\,CNP-} + \int_{0}^{V_{ccrit}} LFN_{Near\,CNP+} + \int_{V_{ccrit}}^{V_{cs}} LFN_{Away\,CNP+}$$

$$-V_{ccrit} \leq V_{cd} \leq 0 \leq V_{ccrit} \leq V_{cs} \rightarrow \int_{V_{cd}}^{V_{cs}} LFN = \int_{V_{cd}}^{0} LFN_{Near\,CNP-} + \int_{0}^{V_{ccrit}} LFN_{Near\,CNP+} + \int_{V_{ccrit}}^{V_{cs}} LFN_{Away\,CNP+}$$

(Eq.A33)

and

$$0 \le V_{cs} \le V_{cd} \le V_{ccrit} \rightarrow \int_{V_{cs}}^{V_{cd}} LFN = \int_{V_{cs}}^{V_{cd}} LFN_{Near\,CNP+}, \quad 0 \le V_{ccrit} \le V_{cs} \le V_{cd} \rightarrow \int_{V_{cs}}^{V_{cd}} LFN = \int_{V_{cs}}^{V_{cd}} LFN_{Away\,CNP+}$$

$$0 \ge V_{cd} \ge V_{cs} \ge -V_{ccrit} \rightarrow \int_{V_{cs}}^{V_{cd}} LFN = \int_{V_{cs}}^{V_{cd}} LFN_{Near\,CNP-}, \quad 0 \ge -V_{ccrit} \ge V_{cd} \ge V_{cs} \rightarrow \int_{V_{cs}}^{V_{cd}} LFN = \int_{V_{cs}}^{V_{cd}} LFN_{Away\,CNP-}$$

$$0 \le V_{cs} \le V_{ccrit} \le V_{cd} \rightarrow \int_{V_{cs}}^{V_{cd}} LFN = \int_{V_{cs}}^{V_{ccrit}} LFN_{Near\,CNP+} + \int_{V_{ccrit}}^{V_{cd}} LFN_{Away\,CNP+}$$

$$0 \ge V_{cd} \ge -V_{ccrit} \ge V_{cs} \rightarrow \int_{V_{cs}}^{V_{cd}} LFN = \int_{V_{cs}}^{-V_{ccrit}} LFN_{Away\,CNP-} + \int_{-V_{ccrit}}^{V_{cd}} LFN_{Near\,CNP-}$$

$$-V_{ccrit} \le V_{cs} \le 0 \le V_{cd} \le V_{ccrit} \rightarrow \int_{V_{cs}}^{V_{cd}} LFN = \int_{V_{cs}}^{0} LFN_{Near\,CNP-} + \int_{0}^{V_{cd}} LFN_{Near\,CNP+}$$

$$V_{cs} \le -V_{ccrit} \le 0 \le V_{cd} \le V_{ccrit} \rightarrow \int_{V_{cs}}^{V_{cd}} LFN = \int_{V_{cs}}^{-V_{ccrit}} LFN_{Away\,CNP-} + \int_{-V_{ccrit}}^{0} LFN_{Near\,CNP-} + \int_{0}^{V_{cd}} LFN_{Near\,CNP+}$$

$$V_{cs} \le -V_{ccrit} \le 0 \le V_{ccrit} \le V_{cd} \rightarrow \int_{V_{cs}}^{V_{cd}} LFN = \int_{V_{cs}}^{-V_{ccrit}} LFN_{Away\,CNP-} + \int_{-V_{ccrit}}^{0} LFN_{Near\,CNP-} + \int_{0}^{V_{ccrit}} LFN_{Near\,CNP+} + \int_{V_{ccrit}}^{V_{cd}} LFN_{Away\,CNP+}$$

$$-V_{ccrit} \le V_{cs} \le 0 \le V_{ccrit} \le V_{cd} \rightarrow \int_{V_{cs}}^{V_{cd}} LFN = \int_{V_{cs}}^{0} LFN_{Near\,CNP-} + \int_{0}^{V_{ccrit}} LFN_{Near\,CNP+} + \int_{V_{ccrit}}^{V_{cd}} LFN_{Away\,CNP+}$$

<div align="center">(Eq.A34)</div>

Eqn (A33) refers to the case of $V_{DS}>0$ where $V_{cs}>V_{cd}$ while eqn (A34) refers to the case of $V_{DS}<0$ where $V_{cd}<V_{cs}$. In each case of the two, there are many subcases that cover all the possible situations. The case for the negative $V_{DS}$ of eqn (A34) is identical with the one of positive of $V_{DS}$ eqn (A33) where $V_{cs}$, $V_{cd}$ have the exact opposite operation since $V_{cd}>V_{cs}$ instead of $V_{cs}>V_{cd}$.

**F. Supplementary Information: IV plots for the rest of the devices similarly to Fig. 3 of the manuscript**

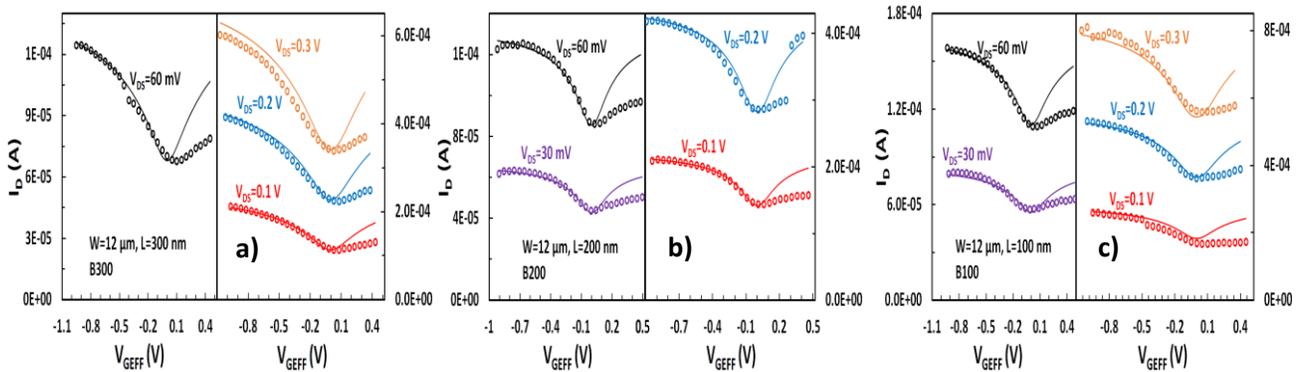

**Fig. S3** Drain current $I_D$ vs. back gate voltage overdrive $V_{GEFF}$, for GFETs with $W=12\ \mu m$ and a) $L=300\ nm$ (B300), b) $L=200\ nm$ (B200) and c) $L=100\ nm$ (B100) at low (left subplot) and high (right subplot) available $V_{DS}$ values ($V_{DS}=30\ mV$, $60\ mV$, $0.1\ V$, $0.2\ V$ and $0.3\ V$). Markers: measured, solid lines: model.

## G. Supplementary Information: *LFN* plots for the rest of the devices similarly to Fig. 6 of the manuscript

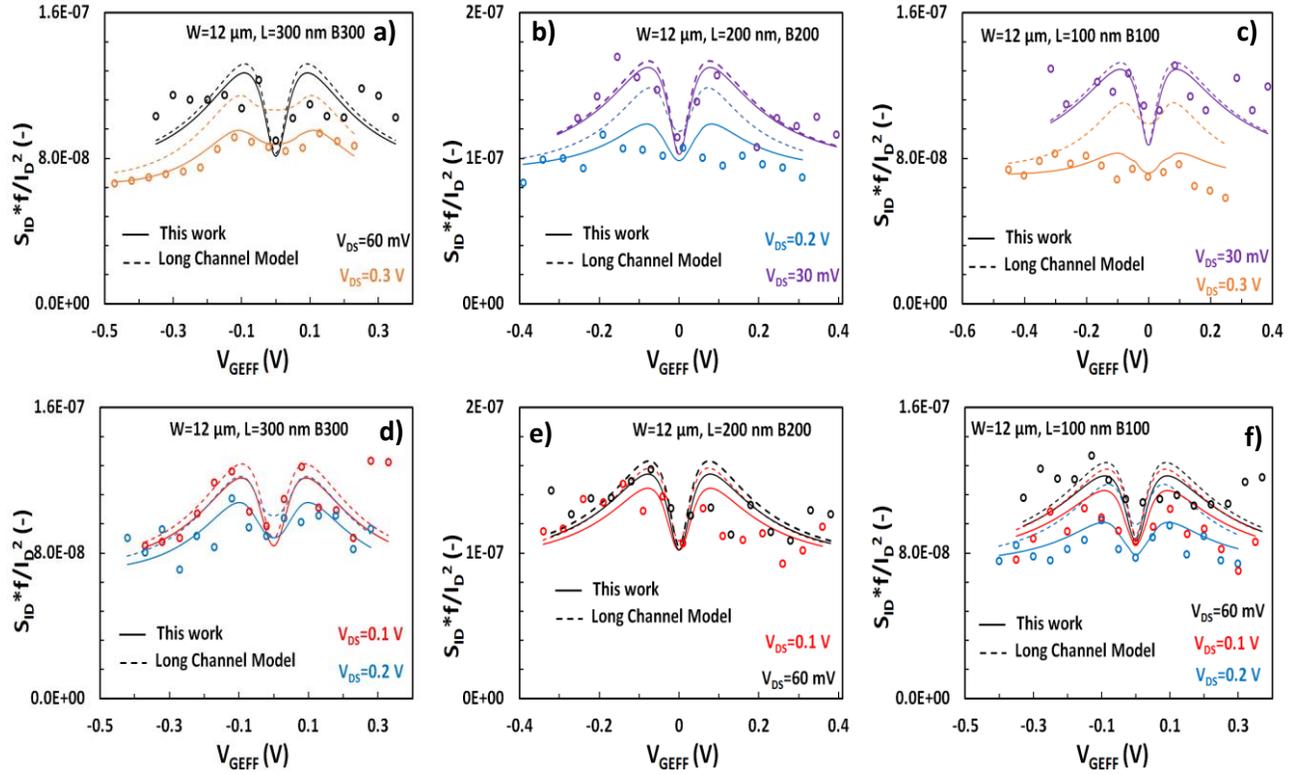

**Fig. S4** Normalized drain current noise divided by squared drain current referred to *1 Hz*, $S_{ID}f/I_D^2$, vs. back gate voltage overdrive $V_{GEFF}$, for GFETs with *W=12 μm* and a, d) *L=300 nm* (B300), b, e) *L=200 nm* (B200) and c, f) *L=100 nm* (B100). Upper plots show the highest and lowest available $V_{DS}$ value depending on the GFET. (B300: $V_{DS}$=60 mV, 0.3 V, B200: $V_{DS}$=30 mV, 0.2 V, B100: $V_{DS}$=30 mV, 0.3 V) while down plots show the rest of available $V_{DS}$ values (B300: $V_{DS}$=0.1 V, 0.2 V, B200: $V_{DS}$=60 mV, 0.1 V, B100: $V_{DS}$=60 mV, 0.1 V, 0.2 V). Markers: measured, solid lines: model, dashed lines: long channel model from ref 35.